\documentclass[fleqn,usenatbib,useAMS]{mnras}

\usepackage{graphicx}	
\usepackage{amsmath}	
\usepackage{amssymb}	
\usepackage{multicol}        
\usepackage{bm}		
\usepackage{pdflscape}	
\usepackage{hyperref}
\usepackage{esdiff}
\usepackage{natbib} 
\usepackage[outdir=./]{epstopdf}
\usepackage[font=scriptsize]{caption}

\usepackage[T1]{fontenc}
\usepackage{ae,aecompl}

\usepackage{newtxtext,newtxmath}

\title[Thermal Conduction in Clumpy Disks]{Thermal Conduction in Clumpy Disks and BLR clouds}

\author[H. Ayad et al.]{
Hussein Ayad $^{1}$,
Maryam Samadi$^{1}$,
Shahram Abbassi $^{1}$ \thanks{abbassi@um.ac.ir}
\\
$^{1}$Department of Physics, Faculty of Science, Ferdowsi University of Mashhad, Mashhad, 91775-1436, Iran\\
}
\date{}

\pubyear{2022}

\begin{document}
\label{firstpage}
\pagerange{\pageref{firstpage}--\pageref{lastpage}}
\maketitle

\begin{abstract}

We investigate the dynamics of clumps that coexisted with/in advection-dominated accretion flows by considering thermal conductivity. Thermal conduction can be one of the effective factors in the energy transportation of ADAFs; hence it may indirectly affect the dynamics of clumps by means of a contact force between them and their host medium. We first study the ensemble of clumps by assuming them as collision-less particles and secondly we find the orbital motion of these clouds as individuals. For both parts, clumps are subject to the gravity of the central object and a drag force. The strong coupling between clumps and ADAF leads to equality between the average treatment of the clumps and the dynamics of their background. By employing the collision-less Boltzmann equation we calculate the velocity dispersion of the clumps which turns out approximately one order of magnitude higher than the ADAF.  In fact, involving drag force in such a system causes the angular momentum of the clumps can be transported outwards by the ADAF, and hence the clouds eventually will be captured at the tidal radius. The results show that the presence of thermal conduction increases the root of the averaged radial velocity square and this in turn speeds up the process of capturing the clouds through the tidal force. In the end, we focus on a typical individual cloud, the spiral orbits appear thanks to only the toroidal component of friction force. The parametric study again proves that the operation of thermal conduction helps for decreasing the lifetime of clumps.

\end{abstract}




\begingroup
\let\clearpage\relax

\section{Introduction}
\label{sec:intro}
Nowadays, we have a lot of valuable information at different wavelengths which enable us to study visible celestial objects like planets, stars, and galaxies as long as some non-visible active astronomical objects such as black holes and neutron stars. It is widely believed that the main engine of these invisible objects is the accretion process. In fact, accretion disks include a wide range of scales: millions of kms in low Mass X-ray Binaries and Cataclysmic Variables, solar radius-to-AU scale disks in protostellar objects, and AU-to-parsec scale disks in Active Galactic Nuclei (Spruit 2010).

Since 1952 that the simplest model for accreting systems with spherical symmetry was proposed by Bondi, and several models of accretion have been introduced to describe the basic features of emergent spectrum arising from the rotating medium around compact objects. The main types of black hole accretion flows are classified based on their apparent shapes and the transportation mechanism of viscously dissipated energy (Yuan \& Narayan, 2014, Narayan et al. 1998; Frank et al. 1992). In the earliest class, the flow forms a geometrically thin (but optically thick) disk with relatively low temperature and the internal pressure support is small (e.g. Frank et al. 1992). In the standard thin disk model, ions and electrons have equal temperatures ($T_i$ = $T_e$), and also the heating and cooling rates are balanced ($Q^- =Q^+$). In the second class, flow is specified with two temperatures, one for ions and an equal or smaller one for electrons. These flows are thermally unstable although both rates of cooling and heating are equal (Piran 1978; cf. Rees et al. 1982). The third class is a slim disk or optically thick hot accretion flow that occurs when the cooling rate becomes less than the heating rate ($Q^-< Q^+$)  and ion temperature becomes equal to electron temperature ($T_i$ = $T_e$). The other important class is advection-dominated accretion flow (ADAF) which is mainly optically thin. Like the second class, the ions' temperature differs from the electrons in this solution. The main factor of this type which remarks it is the photon trapping and carrying viscously dissipated energy towards the disk's center. This inner movement of energy makes the flow stable (Narayan \& Yi 1995). Finally, the model of luminous hot accretion flow is another solution (proposed by Yuan 2003) that the cooling process provides by radiation and decrease of entropy.

In all models mentioned above, the fluid is assumed to be continuous and homogeneous but this is just valid in the limit of low mass accretion rates. In luminous hot accretion flows and slim disks, the mass accretion rate is large even larger than the critical value, so it causes thermal instability and leads to the form of cold dense clumps and hence multiphase accretion flow (Yuan 2003). The existence of such cold clouds embedded in hot accretion disks is now confirmed as the broad-line region (BLR) which successfully explains some of the observational evidence. Gillessen et al. 2012; Burkert et al. 2012 have shown that even near the Galactic center, there are some clouds in orbit around the central supermassive black hole. On the other hand, some numerical simulation clearly shows the formation of cold, dense clumps, likely due to thin-shell instabilities in the shocks formed by the stellar winds (e.g., Vishniac 1994).  Our Galactic center, Sgr A*, is surrounded by young, massive stars. Some of these stars are in the Wolf-Rayet phase and have mass-loss rates that could be as high as $10^{-4}M_\odot yr^{-1}$.  One of these clumps could correspond to the G2 cloud which was discovered on its way to being tidally sheared by Sgr A* (Gillessen et al. 2012, 2013). Furthermore, there is strong evidence from both theoretical and observational works that imply accretion onto some black hole X-ray binaries (BHXB) (Malzac \& Celotti 2002; Merloni et al. 2006), active galactic nuclei (AGN)  (Kuncic et al. 1996; Kumar 1999) including low luminosity AGNs (Celotti \& Rees 1999) and ultra-luminous X-ray sources are clumpy rather than homogeneously continuous. Several instabilities are driven by thermal (Krolik 1998), magneto-rotational (Blaes \& Socrates 2001; 2003), and photon bubble (Gammie 1998) have been reported in accreting systems. In the innermost region of BHXBs, the hard X-ray includes broad iron lines and thermal radiation which are unusual for such a hot fully ionized gas. This strange spectrum can be explained by the existence of cold clumps (Yu et al. 2018). Recent infrared observations have identified low-mass gas clouds, G1 and G2, moving through this exact region. Measuring their interaction with the background gas could therefore provide important information about black hole accretion physics (e.g. Narayan, Ozel \& Sironi 2012).

Current attempts to study these clumpy systems concentrate on answering these main aspects: Understanding the processes that may lead to the formation of BLR clouds is an active research field (e.g., Fromerth \& Melia 2001; Pittard et al. 2003). Moreover, there are noticeable uncertainties about the stability of these clouds and the confinement mechanisms in the light of theoretical arguments and recent numerical simulations (Rees 1987; Krause, Schartmann \& Burkert 2012; Namekata, Umemura \& Hasegawa 2014). Regardless of any current uncertainty over the nature of the confinement mechanisms, the orbital motion of BLR clouds and their radiated emission allow us to estimate the mass of the central black hole (e.g., Marconi et al. 2008; Netzer \& Marziani 2010).

So far, the spectral properties of clumpy accretion flows are investigated by some authors like Guilbert \& Rees (1988),
Malzac \& Celotti (2002), Merloni et al. (2006) and Krolik (1998). The dynamics of cold clumps inside ADAFs between the tidal disruption radius and the transition radius are studied by Wang et al. (2012) (WCL12). They could find analytic solutions for both strong and weak coupling between clumps and ADAF. Ghayuri (2012) studied the broad-line regions by using Jean's equations in cylindrical coordinates. He obtained the distributions of cold clumps that can be found in three configurations of non-disk, disk-wind, and pure disk. Khajenabi et al. (2014) followed WCL12 and took into account the magnetic field effects on clumps' dynamics due to clumps' strong coupling to their ambient medium. In fact, the magnetic field is very important to keep clumps close to the central region of AGNs (Rees 1987).

One of the main ingredients of previous theoretical (or simulation) studies is the true nature of the intercloud medium. The complete physics of these clouds is highly complex and involves several physical processes such as pressure, radiative, centrifugal, gravitational, and magnetic forces; consequently, our knowledge about the intercloud medium is still poor. Most of the previous analytical studies of BLR clouds are based on a few certain simplifying approximations. Recently, it has been suggested by Krause et al. (2011) that one of the plausible candidates to describe the intercloud medium is Advection-Dominated Accretion Flows (ADAFs), where the pressure of the gas varies in proportion to a power-law function of the radial distance (e.g., Narayan \& Yi 1994). Das \& Sharma 2013; Bu \& Gan 2018 have found that cold clumps can co-exist with hot accretion flow. Simulation performed by  Bu \& Gan 2018 has shown that the accretion rate increases to a value at which the radiative cooling rate is roughly equal to or slightly larger than the viscous heating rate, and cold clumps can form around the equatorial plane. On the other hand, Burkert et al. (2012) performed numerical simulations of a clumpy cloud with properties similar to G2 which is moving within the ambient gaseous medium and is modeled as an Advection-Dominated Accretion Flow. they found that despite the variations of the cloud due to its interactions with the ambient gaseous medium the cloud preserves its pressure equilibrium with the surrounding medium (also see, Schartmann et al. 2012). Cloud stability and confinement require them to be in rough pressure equilibrium with their environment (e.g. Krolik, McKee \& Tarter 1981; Krolik 1988). To reach the required pressure of $p \approx10^{-2} dyne/cm^{2}$, the intercloud medium needs either a high temperature ($\approx 10^7 K$; Krolik 1988).  following the above-mentioned papers, here we also describe intercloud medium using an analytical model of ADAFs. In an extremely low accretion rate, the collisional mean free path of electrons is large compared with the length scale of the system, thus thermal conduction can have a significant influence on the dynamics of the accretion flow. On the other hand, in ADAFs, the temperature of the flow is near-virial and the internal energy of particles is very large. Therefore, the extra heat which could not be transported outwards via radiation cooling should be carried through other ways like advection. Besides advection, conduction can transport heating in the radial direction. Tanaka \& Menou (2006) investigated the effect of thermal conduction in ADAFs for the first time. Their work was followed by Shadmehri (2008), Abbassi et al. (2008), and (2010) to find how the dynamics of flow change in the presence of conduction.

In the paper on WCL12, the clumps are treated as stable long-lived objects and because of that, the time dependency of quantities is ignored. However, besides the stationary configuration of clumps, another approach can be followed based on the transient nature of cold clouds. In fact, time evaluation of clumps is an effective tool for finding the population properties of broad-line regions (BLRs) of active galactic nuclei. BLRs are believed to consist of dense clumps
of hot gas in a much hotter diluted medium such as ADAFs and the motion of clouds has the main role in broadening their emission lines. Moreover, the existence of clumps in BLR has been proved by spect-polarimetric observations, too (Smith et al. 2005, Ghayuri 2016).

In order to obtain precious information about the structure of AGN and provide an estimation for the mass of the central
black hole (e.g., Marconi et al. 2008; Netzer \& Marziani 2010), orbital analysis of BLR clouds has attracted some author's attention (Krause, Burkert \& Schartmann 2011; Plewa, Schartmann \& Burkert 2013; Shadmehri 2015; Khajenabi 2015, 2016). To study BLR clouds, it is common to assume the clumps have balanced interior pressure with their ambient pressure. Therefore, a cloud with a fixed mass that is subject to the gravity of the central black hole and the friction force can be treated in the format of a classical two-body problem. Some works (e.g. Mittleman \& Jezewski 1982, Mavraganis \& Michalakis 1994,  Humi \& Carter 2002, Shadmehri 2015, and Khajehnabi 2016) have been done by following this approach. In these papers,  a power-law pressure distribution for the inter-clouds medium has been applied. The common result of all these works is that irrespective of initial conditions or the form of friction force orbital decay occurs in such systems because of considering the drag force. Shadmehri (2015) has introduced the time-of-flight as the estimated time for a cloud needs to fall onto the central region. He found that the drag force coefficients are important in the flight time
of a BLR cloud and change it linearly.

Now in this paper, we firstly work on the ensemble of clumps and employ the collision-less Boltzman equations to describe their dynamics. In section 2, we write the basic equations in cylindrical coordinates and simplify them by assuming ignorable vertical movement for both clumps and the ambient medium. As we know (from WCL12) in the case of strong coupling, the averaged properties of clumps follow the ADAF dynamics. Therefore, both the average velocity of clumps and drag force will be affected by thermal conductivity in the inter-cloud gas. That is why we pay special attention to the characteristics of an ADAF possessing thermal conductivity in section 3. The result of sections 2 and three are combined in section 4 for finding the velocity dispersion between the clumps and the ADAF.
  For the second part of this work and in section 5, we study the dynamics of clumps individually just like the two-body problem and write the momentum equation for two directions (radial and azimuthal) at the equatorial plane. We will use the same formula for drag force as the first part and see how much the trajectory of a BLR cloud changes by adding thermal conduction. And finally, the summary and conclusion are presented in section 6.

\section{Basic Equations and Assumptions}
 As we stated in the introduction existence of cold clouds embedded in hot accretion disks is now confirmed as the broad-line region (BLR) which successfully explains the observational evidence. The formation of cold clouds can be simply justified by thermal instability (TI). Although the detailed calculation of thermal instability is beyond the scope of this paper we may use simplified arguments to grasp some essential properties of cold clouds. The Maximum size of the cold clump is specified by crossing the distance of a sound wave within one Keplerian timescale. Typical values for mass and radius of clump $m_{cl}=4 \times10^{24} g$ and $R_{cl}= 10^{11} cm$, in the case we have a supermassive central black hole with $10^8 M_{\sun}$(Table 1, WCL12). We may expect that the size and mass of the clumps change with the radial distances to the black hole.  On the other hand, the energy transport (cooling and heating of the clump) will specify the minimum size of the clump, below which the clump will be evaporated (WCL12). Furthermore, Sutherland \& Dopita 1993 have shown that due to efficient line cooling clump temperature could retain a constant value around $T=10^4 K$.

One more concern is whether this clump can remain unchanged and survive in an ADAF. Actually, Turbulence in the accretion disk can not destroy these clumps.  MRI- or any other mechanism for turbulence in the disks is only a small disturbance to the clump since the clump size is much smaller than the typical length of turbulence. Moreover, the small turbulence eddies whose size is comparable to the clump have very small kinetic energy according to Kolmogorov's law (Landau \& Lifshitz). Considering all of these arguments it would be an acceptable approximation that clumps are simplified as particles. Furthermore, if we assumed a specific volume for any individual clump we need to take the tidal force into account especially close to the black hole, which may disrupt the clumps. As an approximation for simplifying the complexity of the problem in this study, we assume BLR clouds as a system of point-like particles. So we may ignore the effect of tidal interactions. WCL12 has shown that in Clumpy-ADAF system collision the clumps can be neglected. So it is worth assuming the clumps as a system of collision-less particles were moving inside the ADAFs. Therefore, we need to apply the Boltzmann equation.

Boltzmann equation works with the state of the gas statistically. Firstly we firstly define the distribution function, $ f(\textbf{x},\textbf{v}, t)$ such that $f d^3x d^3v$  which is the average number of particles contained in a volume element $d^3x$
about $x$ and a velocity-space element $d^3v$ about $v$ at time $t$.
Macroscopic properties of the gas such as the number density of the particles, $n$, and the average velocity of an element of gas $u$ (= macroscopic flow velocity) can be evaluated from this function as:
\begin{equation}
n(\textbf{x},t)=\int_{-\infty}^{+\infty}f(\textbf{x},\textbf{v}, t) d^3 v,
\end{equation}
\begin{equation}
u(\textbf{x},t)=n^{-1}\int_{-\infty}^{+\infty}f(\textbf{x},\textbf{v}, t)\textbf{v} d^3 v=\langle v\rangle,
\end{equation}

The distribution function in the absence of collisions is invariant  so we recast the Boltzmann equation as:
\begin{equation}
F=\frac{\partial f}{\partial t}+v^i\frac{\partial f}{\partial x^i}+a^i \frac{\partial f}{\partial v^i}=0
\end{equation}
where $a^i$ is the i-component of the acceleration vector (due to the external force $F(\textbf{x}, t)$, such that  $a(\textbf{x}, t) =F(\textbf{x}, t)/m$ where $m$ is the mass
of a single particle). Equation (3) is known as the collision-less Boltzmann equation, or Vlasov’s equation.
The equations of fluid dynamics can be derived by calculating moments of the Boltzmann
equation for quantities that are conserved in collisions of the particles. Regarding clumps which are assumed to be collision-less, we require the moments of the Boltzmann equation obtained from Eq.(3) and simplifying $\int n F d^3v=0$ that yield:

the zero moments:
\begin{equation}
\frac{\partial n}{\partial t}+\frac{\partial }{\partial x_i}(n\langle v_i\rangle)=0
\end{equation}
and the first moment:
\begin{equation}
\frac{\partial}{\partial t}(n\langle v_i\rangle)+\frac{\partial }{\partial x_i}(n\langle v_i v_j\rangle)-n \langle a_j\rangle=0
\end{equation}
Now converting the Cartesian coordinates to cylindrical coordinates, equations (4) becomes:
\begin{equation}
\frac{\partial n}{\partial t}+\frac{1}{R}\frac{\partial }{\partial R}\big(R n\langle v_R\rangle\big)
+\frac{1}{R}\frac{\partial }{\partial\phi}\big(n \langle v_\phi\rangle\big)+\frac{\partial }{\partial z}\big(n \langle v_z\rangle\big)=0
\end{equation}
and the three components of vectorial Eq.(5) are found as:
 \begin{displaymath}
 \frac{\partial}{\partial t}\big(n\langle v_R\rangle\big)+\frac{1}{R}\frac{\partial}{\partial R}\big(R n\langle v_R^2 \rangle\big)+\frac{1}{R}\frac{\partial}{\partial\phi}\big(n\langle v_R v_\phi\rangle\big)
 \end{displaymath}
\begin{equation}
+\frac{\partial}{\partial z}\big(n\langle v_R v_z\rangle\big)-n\langle a_R\rangle-n\frac{\langle v_\phi^2\rangle}{R}=0,
\end{equation}
\begin{displaymath}
\frac{\partial}{\partial t}\big(n\langle v_\phi\rangle\big)+\frac{1}{R^2}\frac{\partial}{\partial R}\big(R^2 n\langle v_R v_\phi \rangle\big)
\end{displaymath}
\begin{equation}
+\frac{1}{R}\frac{\partial}{\partial\phi}\big(n\langle v_\phi^2\rangle\big)+\frac{\partial}{\partial z}\big(n\langle v_\phi v_z\rangle\big)-n\langle a_\phi\rangle=0,
\end{equation}
\begin{displaymath}
\frac{\partial}{\partial t}\big(n\langle v_z\rangle\big)+\frac{1}{R}\frac{\partial}{\partial R}\big(R n\langle v_R v_z \rangle\big)
\end{displaymath}
\begin{equation}
+\frac{1}{R}\frac{\partial}{\partial\phi}\big(n\langle v_R v_z\rangle\big)+\frac{\partial}{\partial z}\big(n\langle v_z^2\rangle\big)-n \langle a_z\rangle=0,
\end{equation}
after applying assumptions of $\partial/\partial t=\partial/\partial\phi=0$ and also neglecting the terms including: $\langle v_z\rangle$, $\langle v_R v_z\rangle$ and $\langle v_\phi v_z\rangle$, we have:
\begin{equation}
\frac{1}{R}\frac{\partial }{\partial R}\big(R n \langle v_R\rangle\big)=0
\end{equation}

\begin{equation}
\frac{1}{R}\frac{\partial}{\partial R}\big(Rn\langle v_R^2 \rangle\big)
-n\langle a_R\rangle-n\frac{\langle v_\phi^2\rangle}{R}=0,
\end{equation}
\begin{equation}
\frac{1}{R^2}\frac{\partial}{\partial R}\big(R^2 n\langle v_R v_\phi \rangle\big)-n\langle a_\phi\rangle=0,
\end{equation}

\begin{equation}
\frac{\partial}{\partial z}\big(n\langle v_z^2\rangle\big)
-n\langle a_z\rangle=0,
\end{equation}

In equations (11)-(13), we see the acceleration's components, $a_i$, which are specified by two factors: firstly gravitational potential $\psi=-GM/(R^2+z^2)^{1/2}$ and secondly drag force, $D_i=-f_i(v_i-V_i)|v_i-V_i|$ where $V_i$ and $f_i(>0)$ are ADAF velocity component and  coefficient of drag force per unit mass, respectively. Here, it is necessary to determine the sign of drag force:
\begin{equation}
D_i=
\begin{cases}
+f_i(v_i-V_i)^2>0\hspace*{0.5cm}\text{if}\hspace*{0.2cm}v_i<V_i\\
-f_i(v_i-V_i)^2<0\hspace*{0.5cm}\text{if}\hspace*{0.2cm}v_i>V_i\\
\end{cases}
\end{equation}
As we mentioned before, in the current work, we assume that clumps have been strongly coupled with their hot medium. Accordingly, the averaged value of $D_i$ is estimated by applying this approximation: $\langle v_i\rangle=V_i$ and thus the averaged acceleration of drag force can be simplified as:
\begin{equation}
\langle D_i\rangle=- f_i\big(\langle v_i^2\rangle -V_i^2)
\end{equation}
Notice that we use the above formula for both cases of $v_i<V_i$ and $v_i>V_i$. To explain Eq.(15), firstly according to Eq.(14),  $D_i$ can be negative (if $v_i>V_i$) or positive (if $v_i<V_i$), hence it sounds logical not to expect the sign of $\langle D_i\rangle$ becomes different from $D_i$ itself (since in strong coupling, most of clumps must rotate in the same direction of ADAF and moves inwards to join the accretion process) and secondly it seems reasonable to consider $v_i^2<V_i^2$ for the first case and $v_i^2>V_i^2$ for the second case (again because of their similar direction of movements for strong coupling). Consequently, to estimate the approximate value of $\langle D_i\rangle$  we need to note that:
\begin{displaymath}
\langle (v_i-V_i)^2\rangle\approx
\begin{cases}
 V_i^2-\langle v_i^2\rangle\hspace*{0.5cm}\text{if}\hspace*{0.2cm}v_i<V_i\\
\langle v_i^2\rangle-V_i^2\hspace*{0.5cm}\text{if}\hspace*{0.2cm}v_i>V_i\\
\end{cases}
\end{displaymath}
Now we are able to achieve the average value of total acceleration components:
\begin{equation}
\langle a_R\rangle = -\frac{v_K^2}{R} -f_R\big(\langle v_R^2\rangle-V_R^2\big)
\end{equation}
\begin{equation}
\langle a_\phi\rangle=- f_\phi(\langle v_\phi^2\rangle-V_\phi^2)
\end{equation}
\begin{equation}
\langle a_z\rangle =- \frac{z}{R^2} v_K^2-f_z\langle v_z^2\rangle
\end{equation}
where $v_K^2=GM/(R^2+z^2)^{1/2}\approx GM/R^3$ and we have assumed that $V_z=0$, $V_R=\langle v_R\rangle$ and $V_\phi=\langle v_\phi\rangle$.

Let us back to Eq.(10), using the approximation of $\langle v_R\rangle=V_R$, we can find the radius dependency of clumps' density as:
\begin{equation}
n=n_0 \bigg(\frac{R}{R_0}\bigg)^{-3/2}
\end{equation}
where $n_0$ is the density of clumps estimated in the typical radius of $R_0$. Notice that NY94 solved the height-averaged equations of ADAFs, the same result is seen here by integrating the differential Eq.(19) along vertical direction of $z$ and assuming $H\propto R$ or else we would find $n\propto R^{-1/2}$ like Ghayuri (2016, see equation 22 of his paper). The height integrated version of continuity equation yields the constant mass accretion rate for clumps as:
\begin{equation}
\dot{M}_{cl}=-2\pi R H n m_{cl} \langle v_R\rangle
\end{equation}
where $m_{cl}$ is the average mass of individual clumps and $H$ is the half-thickness of medium in $R-$radius which can be approximated by $R$ for an ADAF,  i.e. $H\sim R$. Substituting $\langle a_\phi\rangle$ and $n$ from Eq.(17),(19) in the second equation of momentum and approximating $\langle v_R v_\phi \rangle\approx V_R V_\phi$, Eq.(12) gives:
\begin{equation}
\langle v_\phi^2\rangle=V_\phi\big(V_\phi+\frac{V_R}{2Rf_\phi}\big)
\end{equation}
Since $V_R<0$, we might expect $\langle v_\phi^2\rangle<V_\phi^2$ (see Fig.2, compare two curves marked by dashed and solid lines). In addition, we should notice that  $\langle v_\phi^2\rangle$ becomes smaller towards the center and somewhere it might become zero with a critical value of $f_\phi$ :
\begin{equation}
\Gamma_{\phi c}=-\frac{V_R}{2r_{in}V_\phi}=\frac{3\alpha}{2\sqrt{2\epsilon'(5+2\epsilon')}r_{in}}
\end{equation}
where $\Gamma_{\phi c}=f_{\phi c} R_{Sch}$, ( $R_{Sch}$ is the Schwarzschild radius) and like NY94 for convenience we have defined the parameter of  $\epsilon'$ instead of $(5/3-\gamma)/[f(\gamma-1)]$ and also scaled inner radius as: $r_{in}=R_{in}/R_{Sch}$. 

This critical value of $\Gamma_{\phi }$ is similar to Eq.(26) of WCL12 but independent of the radius and only changes by previously known parameters of this problem, i.e. $\alpha, \gamma,f$ and $r_{in}$. In Fig.1 we have shown $\Gamma_{\phi c}$ as a function of $\gamma$ for different $f$'s and a constant $\alpha (=0.1)$.

\endgroup  

\begin{figure}
\centering
\includegraphics[width=80mm]{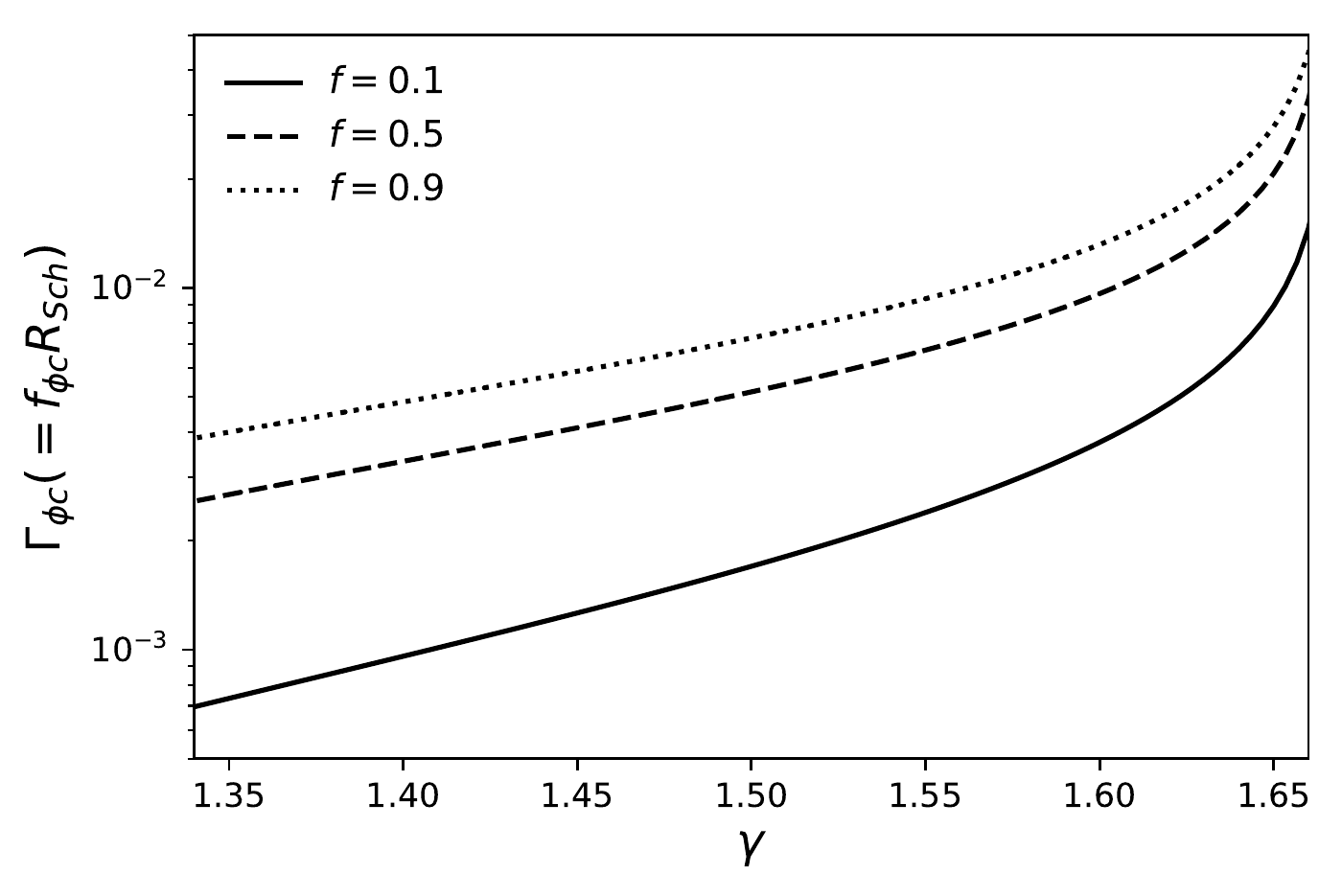}
\caption{ Critical value of rotational drag force coefficient with respect to the polytropic parameter with different $f$'s. }
\end{figure}
\begin{figure}
\centering
\includegraphics[width=70mm]{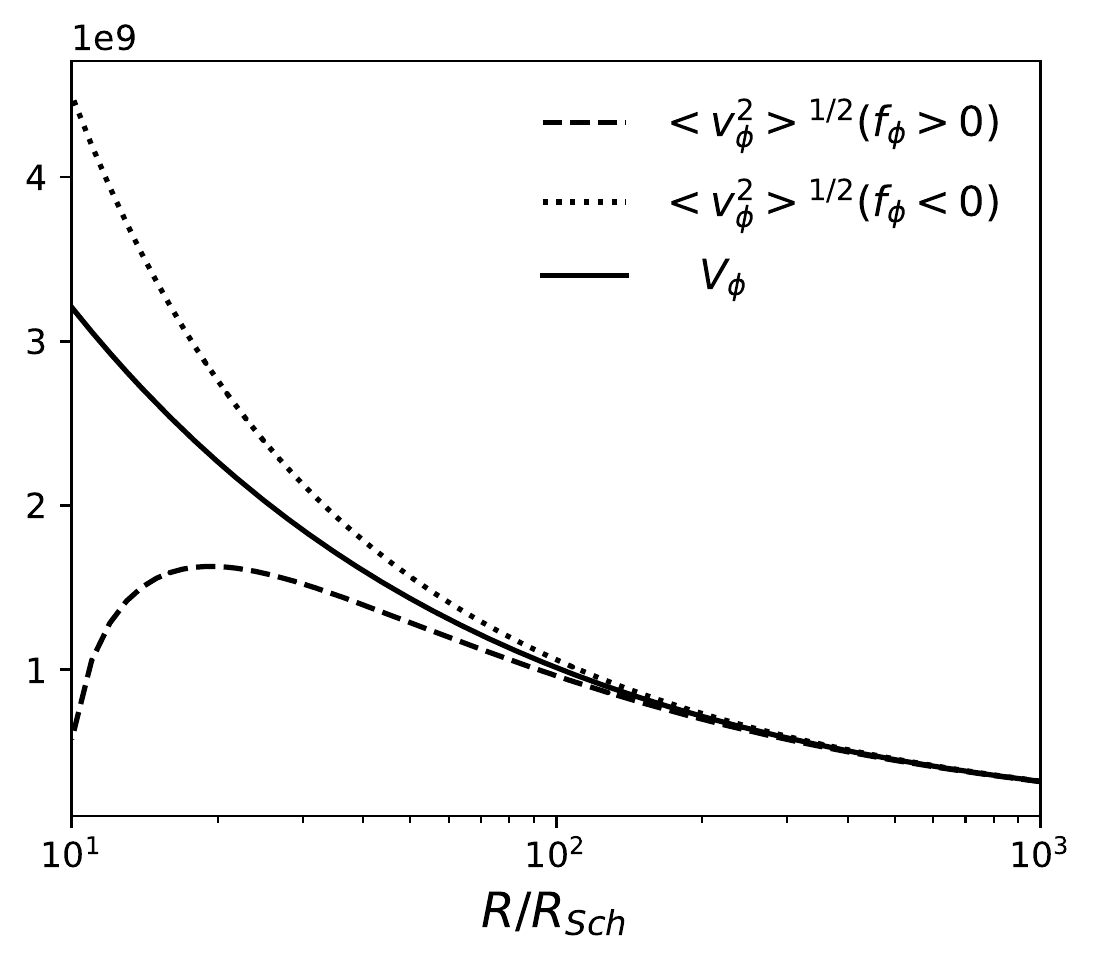}
\caption{ Variation of the rotational velocity squared as a function of radius, $r$. In this figure, the parameters are: $f=0.9, \alpha=0.1, \gamma=1.4$ and $f_\phi R_{Sch}= \Gamma_\phi=\pm 5\times 10^{-3}$, where $R_{Sch}=2.95 \times 10^5 m_*\hspace*{0.1cm} (cm)$ is the Schwarzschild radius.}
\end{figure}
\begin{figure}
\centering
\includegraphics[width=70mm]{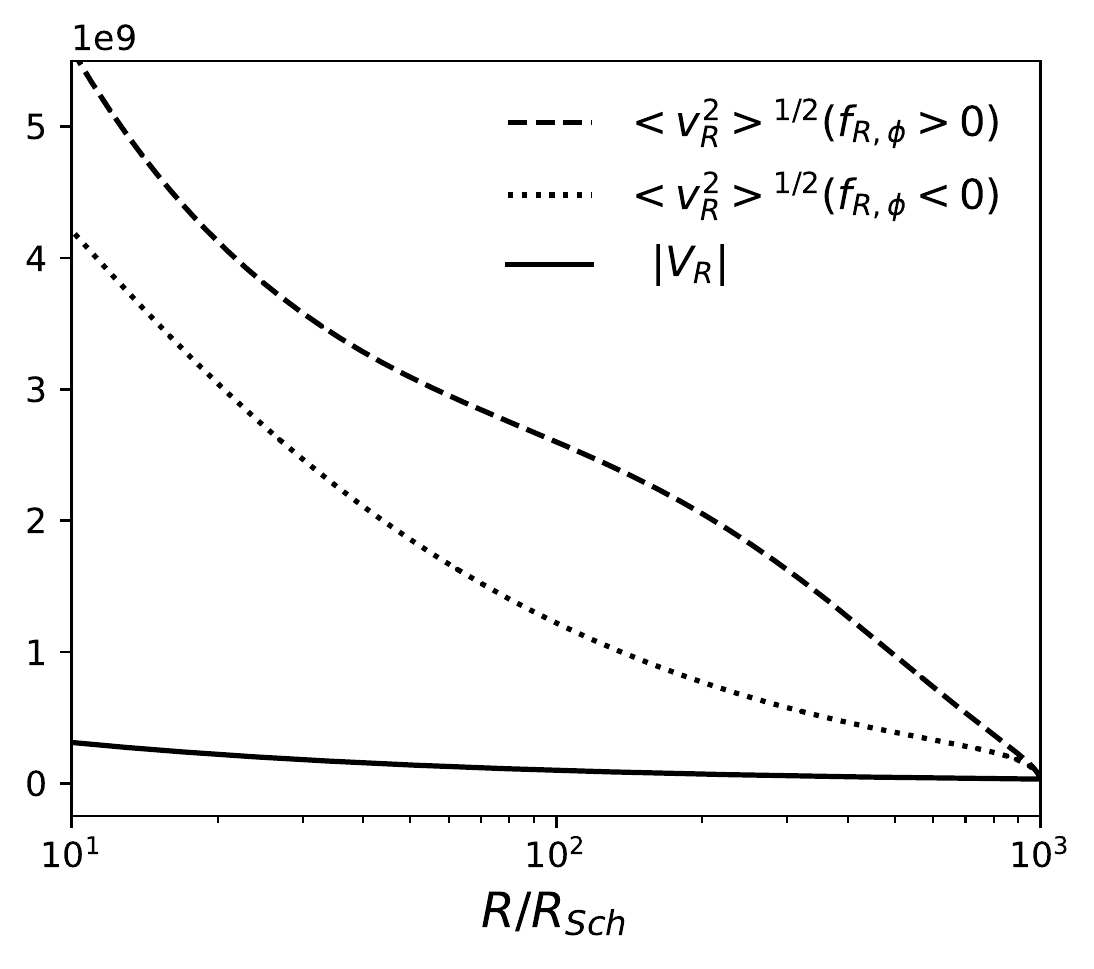}
\caption{ Variation of the radial velocity squared as a function of radius, $R$. In this figure, the parameters are: $f=0.9, \alpha=0.1, \gamma=1.4, f_{R,\phi} R_{Sch}= \Gamma_{R,\phi}=\pm 5\times 10^{-3}$.}
\end{figure}

\begin{figure*}
\centering
\includegraphics[width=175mm]{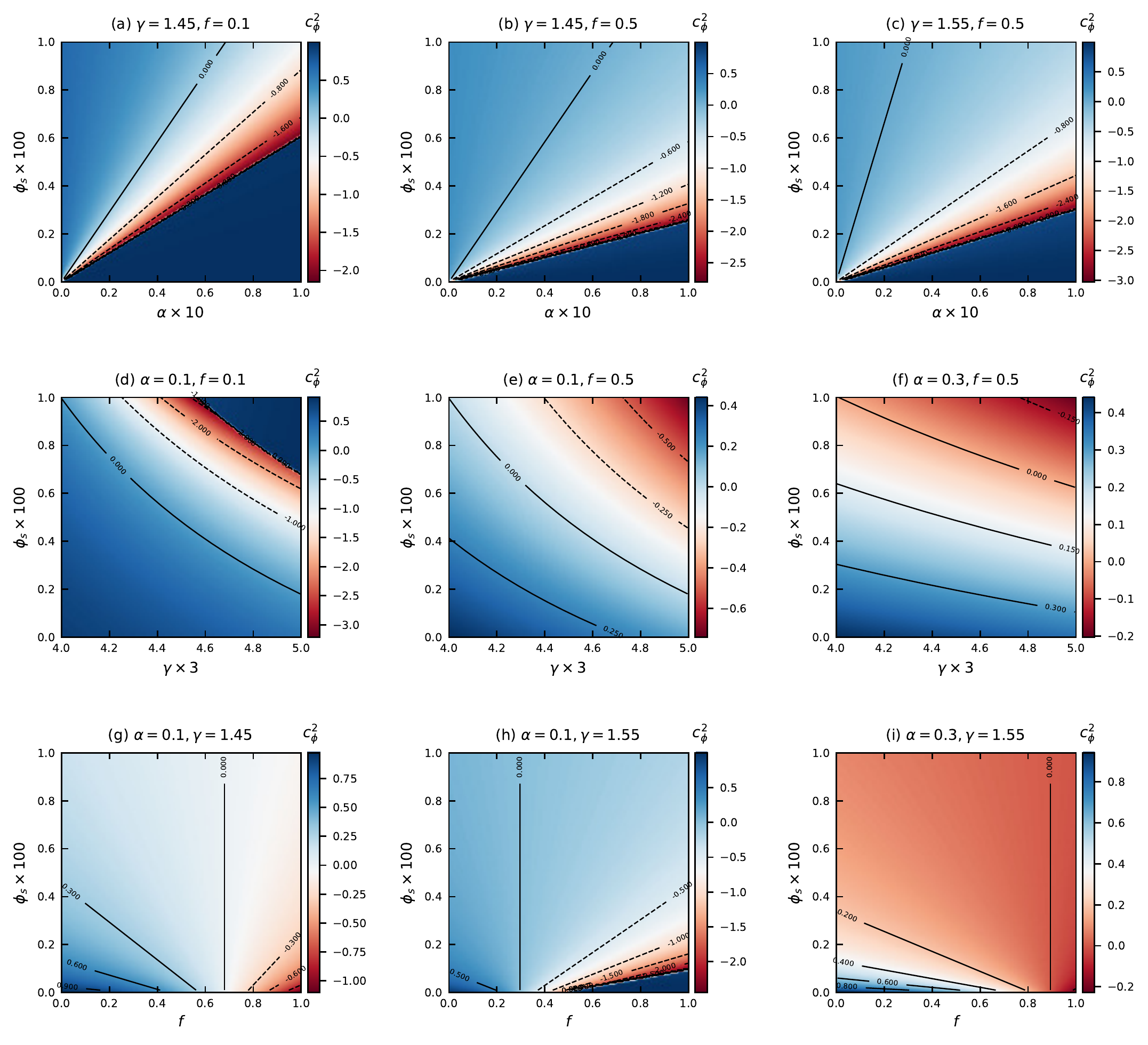}
\caption{Variation of $c_{\phi}^2$ showed by colors with the parameter of thermal conduction $\Phi_s$ and one of three input parameters: $\alpha$ (a-c),$f$ (d-f) , $\gamma$ (g-i). In this figure, the negative values of $c_\phi^2$ are not physical and we should carefully pick up a proper value of $\phi_s$ for each certain set of input parameters $(\alpha, f, \gamma)$ to avoid forbidden areas.}
\end{figure*}

\begingroup

The last achievable quantity here is $\langle v_R^2\rangle$ appeared in Eq.(11). This equation can be simplified by substituting $\langle a_R\rangle$ and $\langle v_\phi^2\rangle$  from Eq.(16) and (21). So we have:
\begin{displaymath}
\frac{d\langle v_R^2\rangle }{d R}-\frac{1}{2R}\langle v_R^2\rangle+\frac{v_K^2}{R}+f_R\big(\langle v_R^2\rangle -V_R^2\big)-\frac{V_\phi}{R}\big(V_\phi+\frac{V_R}{2Rf_\phi}\big)=0
\end{displaymath}
It will be more convenient, if we use $R_{Sch}$ as a length scale and convert $R$ to $r(=R/R_{Sch})$. So by multiplying the above equation by $R_{Sch}$ and applying the definition of $\Gamma_i=f_i R_{Sch}$, we find:
 \begin{equation}
\frac{d\langle v_R^2\rangle }{d r}-\frac{\langle v_R^2\rangle}{2r}+\frac{v_K^2}{r}+\Gamma_R\big(\langle v_R^2\rangle -V_R^2\big)-\frac{V_\phi}{r}\big(V_\phi+\frac{V_R}{2r\Gamma_\phi}\big)=0
\end{equation}
If we write the velocity components of ADAF as: $V_i=c_i v_K(r)$, we can obtain the following result by solving the above equation:
\begin{displaymath}
\langle v_R^2\rangle=c_K^2 r^{1/2}e^{-\Gamma_R r}\bigg\{C_1+  c_R^2 \Gamma_R\int_r^{r_f} r^{-3/2} e^{\Gamma_R r}dr
\end{displaymath}
\begin{equation}
-(1-  c_\phi^2) \int_r^{r_f}  r^{-5/2} e^{\Gamma_R r}dr+ \frac{ c_Rc_\phi }{2\Gamma_\phi}\int_r^{r_f} r^{-7/2}e^{\Gamma_R r} dr  \bigg\}
\end{equation}
where $c_K^2=GM/R_{Sch}(=c^2/2)$ and $C_1$ is a constant value which can be obtained from this boundary condition: $\langle v_R^2\rangle|_{r_f}=V_R^2 |_{r_f}$ ($r_f=1000R_{Sch}$ is the outer boundary where the ADAF solutions are satisfied)  then we find,
\[
C_1=c_R^2 r_f^{-3/2} e^{\Gamma_R r_f}\]
With using $c_R, c_\phi$ and $c_K$, Eq.(21) changes as below:
\begin{equation}
\langle v_\phi^2\rangle=c_\phi\big(c_\phi+\frac{c_R}{2r\Gamma_\phi}\big)c_K^2 r^{-1}
\end{equation}

In the next section, we will find the constants of $c_R$ and $c_\phi$ for an ADAF with thermal conduction.

\section{ADAF solutions with thermal conduction}
We are interested in analyzing the structure of advection-dominated accretion flow (ADAF) where thermal conduction plays an important role in energy transportation. Here, we write the basic conservation equations for an ADAF with the same assumptions as NY94 and involve the thermal conduction term in energy equation (Tanaka \& Menou 2006). We assumed a
steady axisymmetric accretion flow, $\frac{\partial}{\partial t}=\frac{\partial}{\partial \phi}=0$ .we
can write the standard equations in the cylindrical coordinates
($r, \phi,  z$). In addition, we vertically integrate the flow equations
and therefore all the physical variables only become functions
of radial distance r. Moreover, relativistic effects are neglected
and Newtonian gravity in the radial direction is taken into
account. The simplified form of radial and azimuthal components of momentum equation and energy equation are presented in the three following equations:

  \begin{equation}
  V_R \frac{d V_R}{dR}= \frac{V_\phi^2 - v_k^2}{R}-\frac{1}{\rho} \frac{d(\rho C_s^2)}{dR}
  \end{equation}
    \begin{equation}
  V_R \frac{d(RV_\phi)}{dR}= \frac{1}{\rho R H} \frac{d}{dR}\bigg[\frac{\alpha \rho C_s^2 R^4 H}{v_k} \frac{d}{dR}\left(\frac{V_\phi}{R}\right)\bigg]
  \end{equation}
  \begin{equation}
2 H \rho V_R T \frac{ds}{dR}=f \frac{2 \alpha \rho C_s^2 R^3 H}{v_k}\bigg[ \frac{d}{dR}\left(\frac{V_\phi}{R}\right)\bigg]^2 - \frac{2H}{R} \frac{d(R F_s)}{dR}
  \end{equation}
    where $T, s$ and $f$ are the temperature, entropy and advection parameter, respectively; and $H=RC_s/v_K$ is the vertical thickness of hot accretion flow, $F_s = 5 \Phi_s \rho C_s^3$ is the saturated conduction flux (Cowie \& McKee 1977) and $\Phi_s$ is the saturation constant, respectively. According to the self-similar method for the radial direction, all quantities with velocity dimension should be proportional to $R^{-1/2}$ and density is assumed to be proportional to $R^{-3/2}$.
 Consequently, Eq.(26)-(28) give us,

  \begin{equation}
c_\phi^2=1 -  \frac{1}{2} c_R^2  - \frac{5}{2}c_s^2,
  \end{equation}
  \begin{equation}
  c_R = - \frac{3}{2} \alpha c_s^2,
   \end{equation}
   \begin{equation}
   \frac{9}{8}  \alpha^2 c_s^4 + \left[\frac{5}{2} + \frac{5 -3 \gamma}{3f (\gamma-1)}\right]c_s^2  - \frac{40 \Phi_s}{9 \alpha f} c_s  - 1=0
   \end{equation}

\endgroup

\begin{figure*}
\centering
\includegraphics[width=175mm]{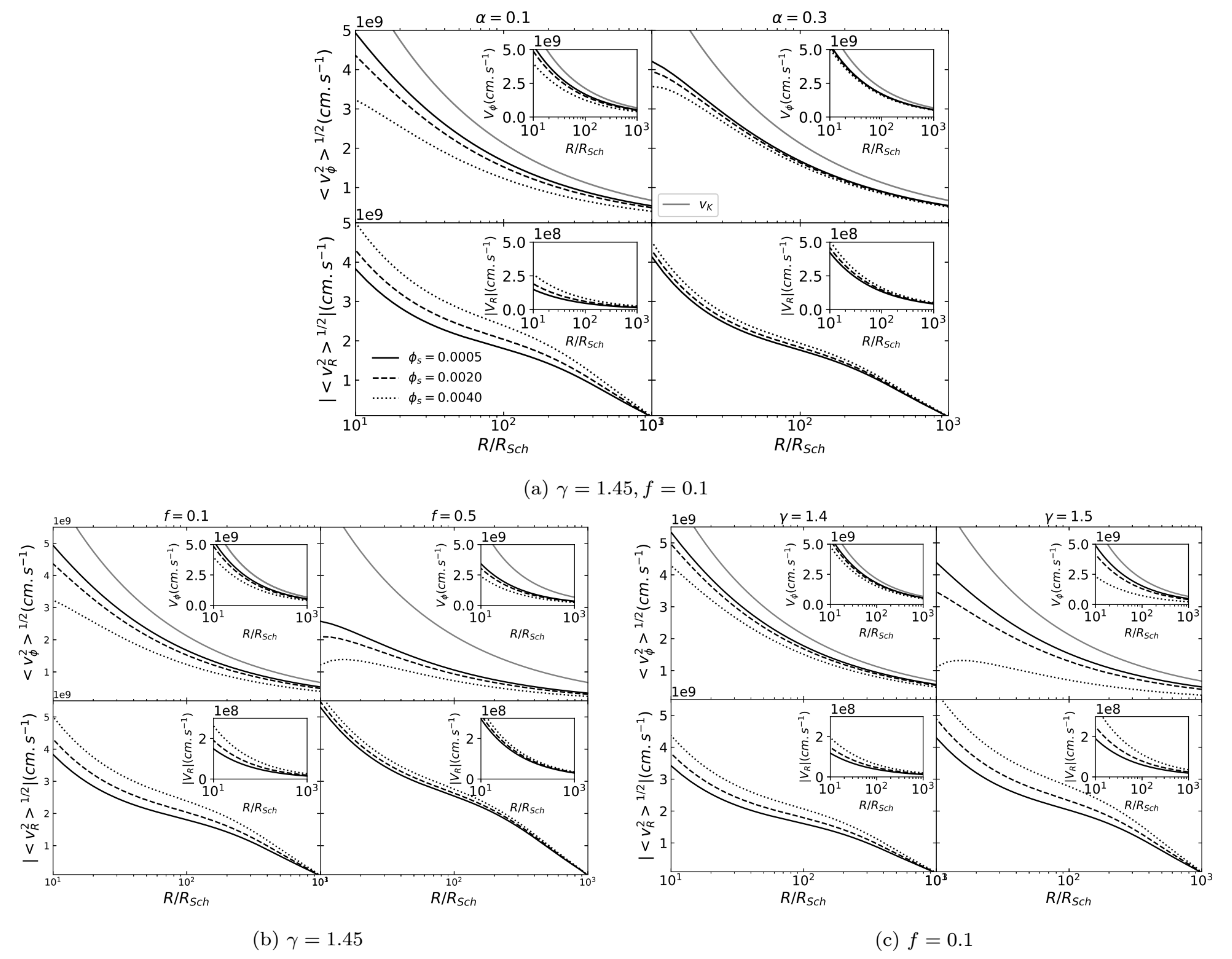}
\caption{Solution of the clumpy-ADAF disk in the presence of thermal conduction: $\Phi_s=0.0005$ (solid line), $\Phi_s=0.0020$ (dashed line), $\Phi_s=0.0040$ (dotted line). The three panels show the dependence of the solution on the three parameters $\alpha$ (panel a), $f$ (panel b), and $\gamma$ (panel c). All of these panels was set up for $\Gamma_{R}=0.005,\Gamma_{\phi}=0.01$.}
\label{fig:fig}
\end{figure*}


where $c_{R,\phi}=V_{R,\phi}/v_K$ and $c_s=C_s/v_K$.   As seen, there are three unknown variables, i.e. $c_R, c_\phi$ and $c_s^2$ in the above equations which can be found numerically. Now we firstly solve the algebraic Eq.(31) in order to find $c_s^2$ and secondly we substitute the solution of $c_s^2$ in Eq.(29) which has been simplified as below by applying Eq.(30) and (31):
\begin{displaymath}
 c_\phi^2=\frac{5 -3 \gamma}{3f (\gamma-1)}c_s^2  - \frac{40 \Phi_s}{9 \alpha f} c_s
\end{displaymath}
Notice that substituting some values of $\Phi_s$ in the solutions yields $c_\phi^2<0$ ( this issue has been discussed in Ghasemnezhad et al. (2012)). This point will be clarified better by looking at the contour plots in figure 4. This figure shows the variation of $c_\phi^2$ with respect to $\Phi_s$ and one of three other input parameters: $\alpha, f, \gamma$ in each subplot. In the first row panels of Fig.4, the unacceptable region for $c_\phi^2$ is located between two radial lines in the $\alpha-\phi_s$ plane and it gets larger as the advection (in panel b) or polytropic parameter (in panel c)  increases.  The second three panels displays contour plots of $c_\phi^2$ in $\gamma-\phi_s$ plane. The boundaries between negative and positive values of rotational velocity squared in the middle row panels are arc-shaped and placed on the right-upper side of panels (d)-(f). Comparing panel (d) with panel (e), a rise in advection causes an upward shift in the forbidden region. On the other hand, in comparison (d) with panel (f), we found out that when viscosity becomes stronger, the curved line of zero rotation seems flattered and moves to higher values of the thermal conduction parameter. Finally, the bottom panels of Fig.4 present the behavior of $c_\phi^2$ by colors with changing $\phi_s$ and $f$. As seen, in each panel there is a vertical line with zero labels which declares that choices of two input parameters of $\phi_s$ and $f$ from the right of this line cannot result in a reasonable solution. Furthermore, the rectangular unacceptable part shrinks by greater viscosity (seen in panel i) but extends by growing $\gamma$. Consequently, in the light of this figure, we can find a proper value of $\phi_s$ for each certain set of input parameters $(\alpha, f, \gamma)$ to avoid forbidden areas.

  In the following section, we will use these obtained solutions of the components of ADAF's velocity for finding drag force and calculating the root of the averaged radial and rotational velocities square.
  

\begin{figure*}
\centering
\includegraphics[width=175mm]{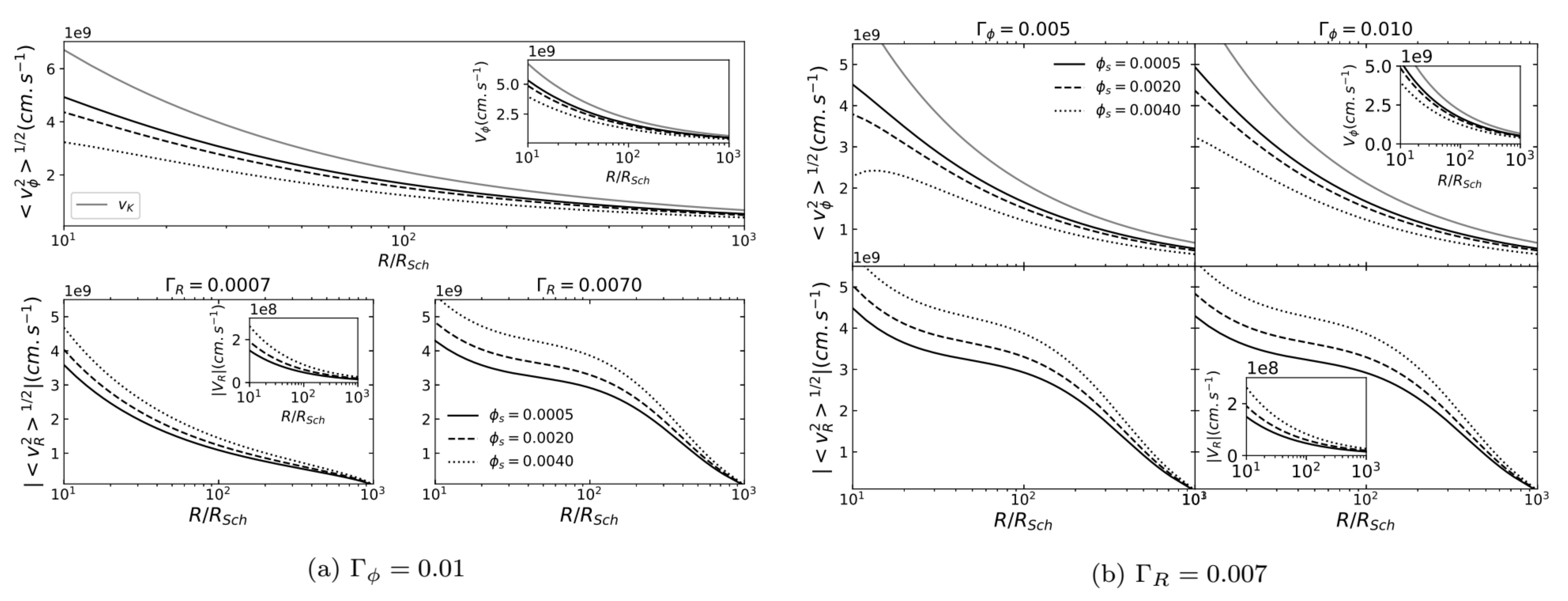}
\caption{Solution of the clumpy-ADAF disk in the presence of thermal conduction: $\Phi_s=0.0005$ (solid line), $\Phi_s=0.0020$ (dashed line), $\Phi_s=0.0040$ (dotted line). the three panels show the dependence of the solution on the three-parameter $\Gamma_{R}$ (panel a) in this panel we take $\Gamma_{\phi}=0.01$, $\Gamma_{\phi}$ (panel b)when $\Gamma_{R}=0.007$ and $\gamma$ (panel c). All of this panels was set up for $\gamma=1.45,f=0.1$.}
\label{fig:fig}
\end{figure*}


\section{The effect of thermal conduction on clumpy disks}

In two previous sections, we reviewed momentum equations of clumpy ADAFs with different signs of drag force and solved the simplified first-order differential equation of $\langle v_R^2\rangle$ with respect to radial coordinate. The final result in Eq.(24) as a function of $c_R$ and $c_\phi$, i.e. coefficients of ADAF's velocity components in radial and azimuthal directions. Now it is time to substitute the numerical ADAF solutions from Eq.(29)-(31) in Eq.(24) and (25). The results are illustrated in figures 5 and 6. In these two figures, we have investigated the influence of thermal conduction on the dynamics of clumps for a different set of input parameters: $(\alpha, f, \gamma)$ in Fig.5, and $(\Gamma_R, \Gamma_\phi)$ in Fig.6 . In these panels, we have considered $\Phi_s=0.0005, 0.002$ and $0.004$, remarked by dotted, dashed and solid lines, respectively. In order to have an easier comparison, both velocity components of ADAF and clumps are presented in Figures 5 and 6. The velocity dispersion of clumps is presented in the main larger plots, and the small plots inside the main plots demonstrate the velocity of ADAF. Comparing the small and large plots declares that thermal conduction plays a similar role in changing the behavior of clumps and ADAF.

The upper row panels of figures 5, 6 show the behavior of ADAFs (or equivalently approximated clumps' azimuthal velocity in small subplots inside the main plots) as well as the average rotational velocity square of clumps (in the bigger main plots). In all upper plots, gray curves show Keplerian velocity and assure us that the velocity dispersion in the azimuthal direction remains sub-Keplerian even after adding $\Phi_s$. In this figure and Fig.6, besides $\Phi_s$, we have also examined two different values of each parameter, and we would like to investigate in which circumstances the effect of thermal conduction on the dynamics of the system becomes more significant. In Fig.5, and 6, we can easily see that the maximum variation of both components of velocity dispersion under the effect of thermal conduction occurs at the inner boundary of the ADAF (or equivalently at the captured radius of clumps) with each of these input parameters: $\alpha$, $\gamma$, and $f$.
Regardless of the value of each parameter, all these plots reveal that an increase of $\Phi_s$ reveals that the rotation of the hot medium will decrease as long as At the same time It is also clear that this change in thermal conduction results in decreasing of the average rotational velocity square of clumps.  Comparing upper right panels with left ones leads us to know that  $\left\langle v_{\phi}^2\right\rangle ^{1/2}$ remarkably decreases with increasing advection parameter and polytropic index but does not behave uniformly with changing the viscosity parameter. Furthermore, according to Fig.6 (two upper right panels), an enhancement of the coefficient of drag force in $\phi$-direction causes a bit increase in this quantity whereas changing of drag force in the radial direction (as we had seen in Eq.25) does not have any direct effect on the average azimuthal velocity square of clumps. One more point to consider, if we look at the numbers of vertical axes of the main plots with ones in small plots, we realize that the velocity dispersion of clumps in the azimuthal direction and rotational velocity of ADAF possess the same order of magnitude.

The lower panels of figures 5, and 6 display directly the role of thermal conduction in a variation of $\left\langle v_{R}^2\right\rangle ^{1/2}$ (presented in main bigger plots) and also in changing of $V_R$ or $\langle v_R\rangle$ (presented in small subplots). According to these plots, we can find out unlike the $\Phi_s$ effect on velocity in the azimuthal direction, the total effect of $\Phi_s$ in the radial direction is positive on $\left\langle v_{R}^2\right\rangle ^{1/2}$ and also on $\left\langle v_{R}\right\rangle$. Here again, we catch our attention to the scale of vertical axes numbers: $10^9$ for main plots and $10^8$ for small subplots, so the radial velocity dispersion of clumps is about 10 times greater than $V_R$ of ADAFs, this result is consistent with WLC12. Moreover, if we compare right and left panels in the bottom rows of each four-subplot shape, we notice that the difference between plots with the various value of $\Phi_s$ is more visible with smaller $\alpha$ and $f$ but bigger $\gamma$ and $\Gamma_R$. Finally, according to the four subplot shapes in the right hand of Fig.6, $\left\langle v_{R}^2\right\rangle ^{1/2}$ is not very sensitive to changing $\Gamma_\phi$ which was pointed out in WLC12 too.


\begin{figure*}
     \centering
\includegraphics[width=\textwidth]{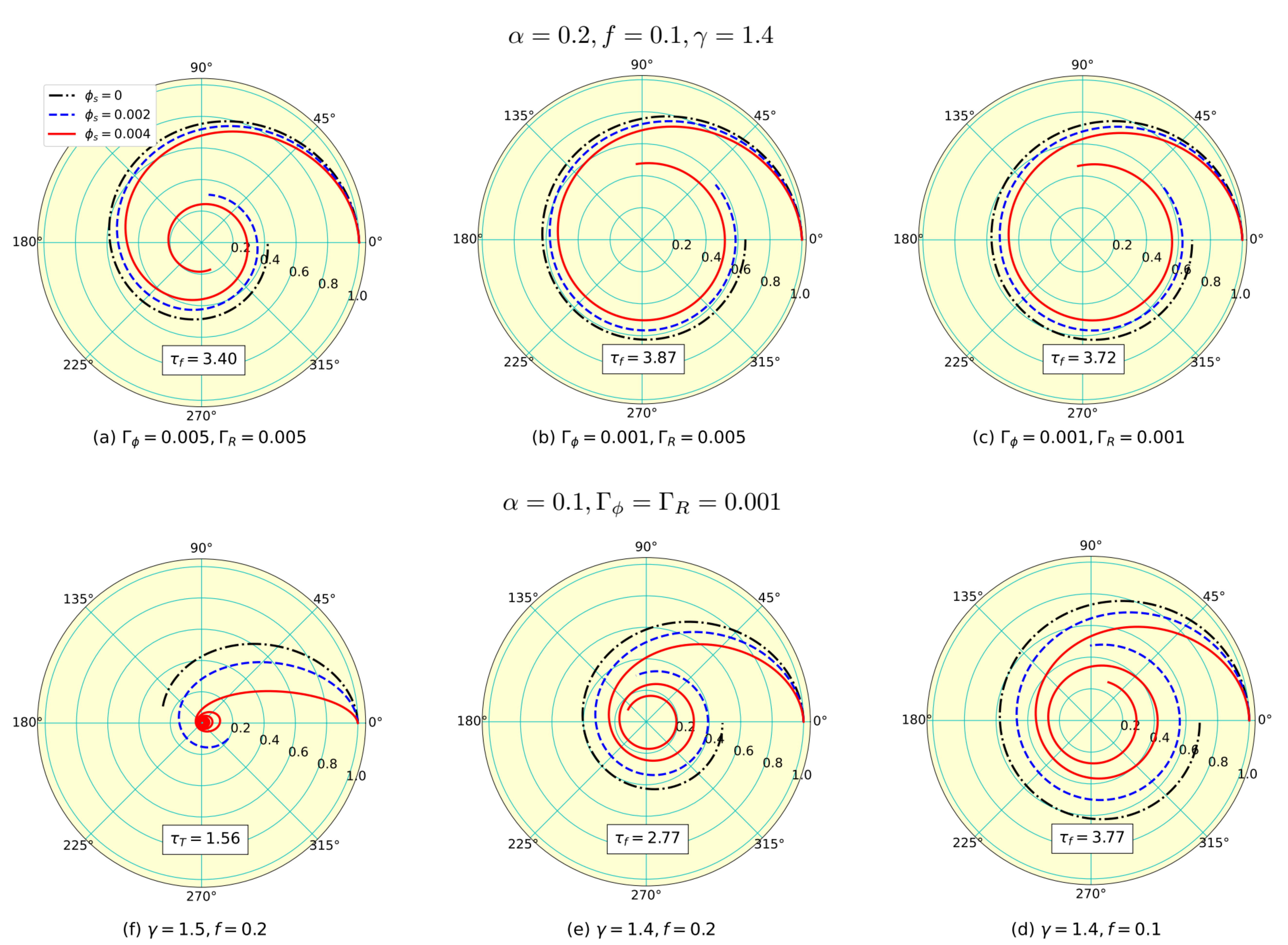}
        \caption{The effect of thermal conduction on the orbital motion of a clump embedded in an ADAF. In this figure, we have neglected the radiation force from the central object, thus the BLR cloud here moves only at the equatorial plane. In each panel, we have fixed the set of input parameters and examined three different values of the thermal conduction parameter, i.e. $\phi_s=0, 0.002, 0.004$. In the upper panels, we have used: $\alpha=0.2, f=0.1, \gamma=1.4$ and in the bottom panels: $\alpha=0.1, \Gamma_R=\Gamma_\phi=0.001$. Initial conditions are: $r|_{t=0}=1$( or $R|_{t=0}=1000 R_{Sch}$), $\phi|_{t=0}=0$, $\dot{r}|_{t=0}=c_R$ (or $\dot{R}|_{t=0}=V_R$) and $\dot{\phi}|_{t=0}=c_\phi$ (or $R\dot{\phi}|_{t=0}=V_\phi$). As it is seen, the trajectory of cloud varies with $\Gamma_\phi$ [compare panel (a) with (b)], $\Gamma_R$ [look at panels (b) and (c)], $\alpha$ [panel (c):$\alpha=0.2$ and panel (d):$\alpha=0.1$] ,$f$ [compare panel (d) with (e)] and $\gamma$ [see panel (e) and (f)]. The final time, $\tau_f$ in panels (a)-(e) is equal to the time required for one round of rotation in the case of $\phi_s=0$, but $\tau_f=1.56$ in panel (f) is the total flight time of cloud to reach the inner boundary of ADAF, or captured radius of clumps in the case of $\phi_s=0.004$. }
 \end{figure*}

 \begin{figure*}
     \centering
  \includegraphics[width=\textwidth]{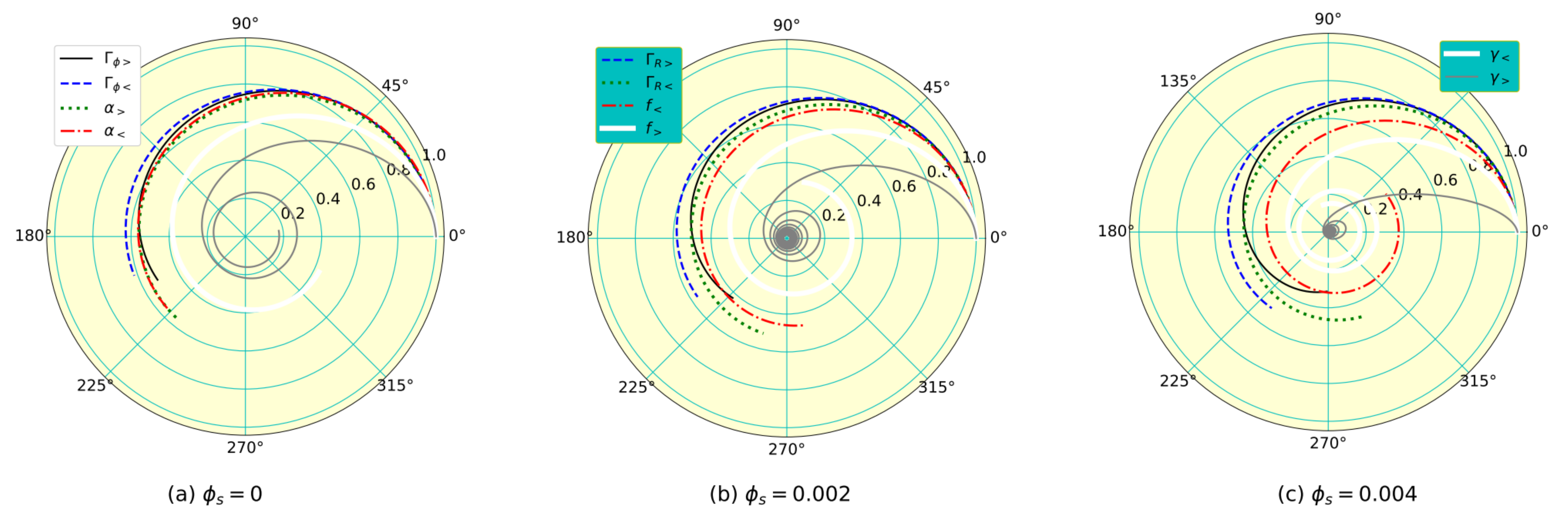}
          \caption{Spiral motion of a BLR cloud in the non-static medium like an ADAF. The input parameters of this figure are similar to figure 7 but we have selected $\tau_f=2.5$ for all curves except for that one in panel (c) with gray color ($\tau_f=1.56)$. In each panel we have used: 1) black solid line for orbital motion by applying the same parameters as  Fig.7a, 2) blue dashed line $\rightarrow$ Fig.7b, 3) green dotted line $\rightarrow$ Fig.7c, 4) red dot-dashed line $\rightarrow$ Fig.7d, 5)white line $\rightarrow$ Fig.7e, 6)gray line $\rightarrow$ Fig.7f. Therefore, 1) black curves have larger $\Gamma_\phi$ comparing with blue dashed ones, 2) green dotted lines have smaller $\Gamma_R$ in comparison with blue ones but it has larger $\alpha$ relative to red dot-dashed curves ($\alpha=0.1$),  and 3) finally white curves have larger $f$ comparing with red ones but smaller $\gamma$ if we compare them with gray curves ($\gamma=1.5$).    }
        \label{fig:three graphs}
     \end{figure*}

\begin{figure*}
\centering
\includegraphics[width=160mm]{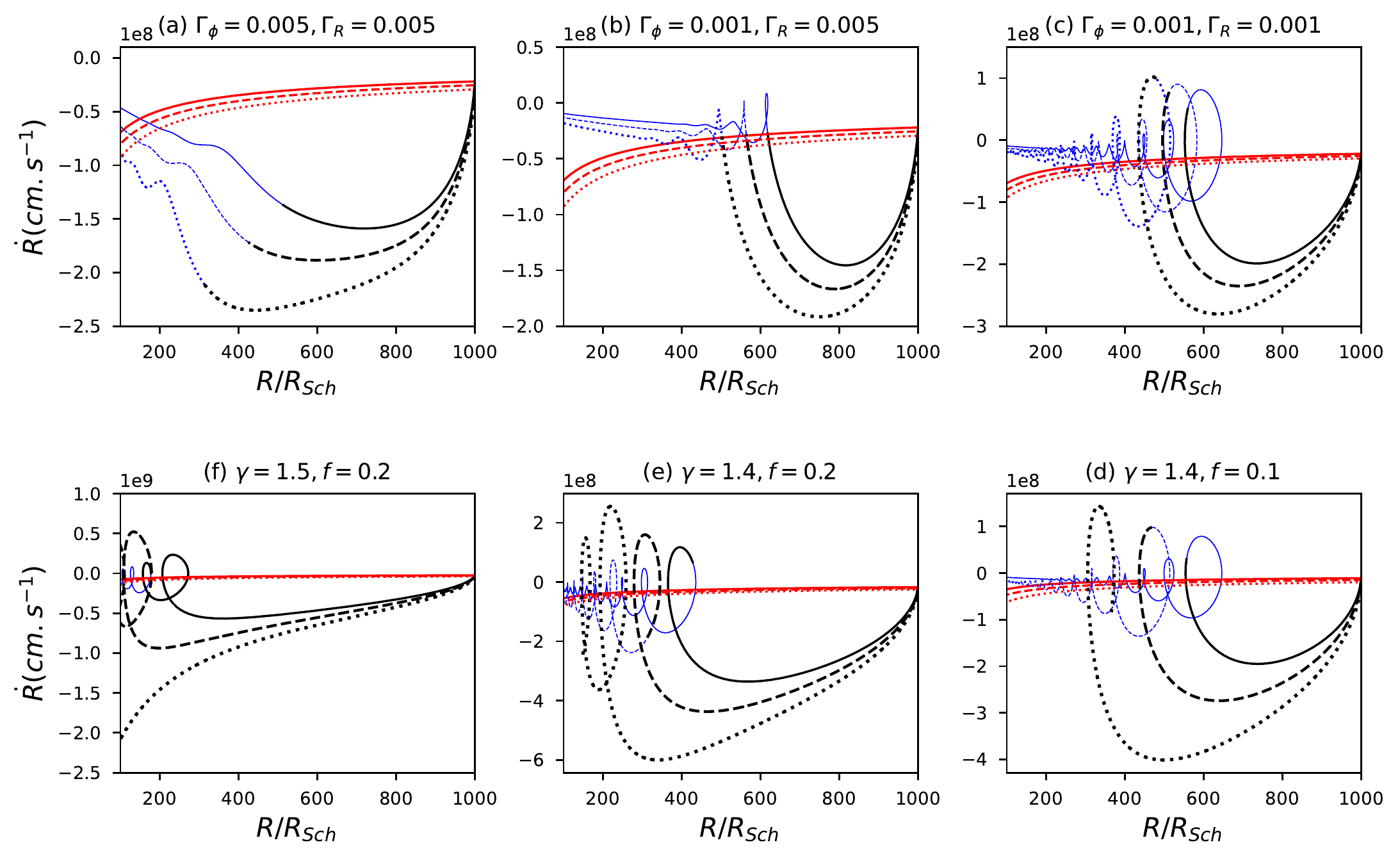}
\caption{Time-dependent variation of the radial velocity of a BLR cloud with respect to radius. Input parameters of each panel are the same as the ones in figure 7, so comparing every pair of adjacent subplots yields information about the effect of a certain parameter other than $\phi_s$. In each panel, the solid dashed and dotted lines refer to solutions with $\phi_s=0, 0.002, 0.004$, respectively. Two colors have been used in this figure; the black curves represent $\dot{R}$ and the red ones show $V_R$ of the ADAF.}
\end{figure*}
\begin{figure*}
\centering
\includegraphics[width=160mm]{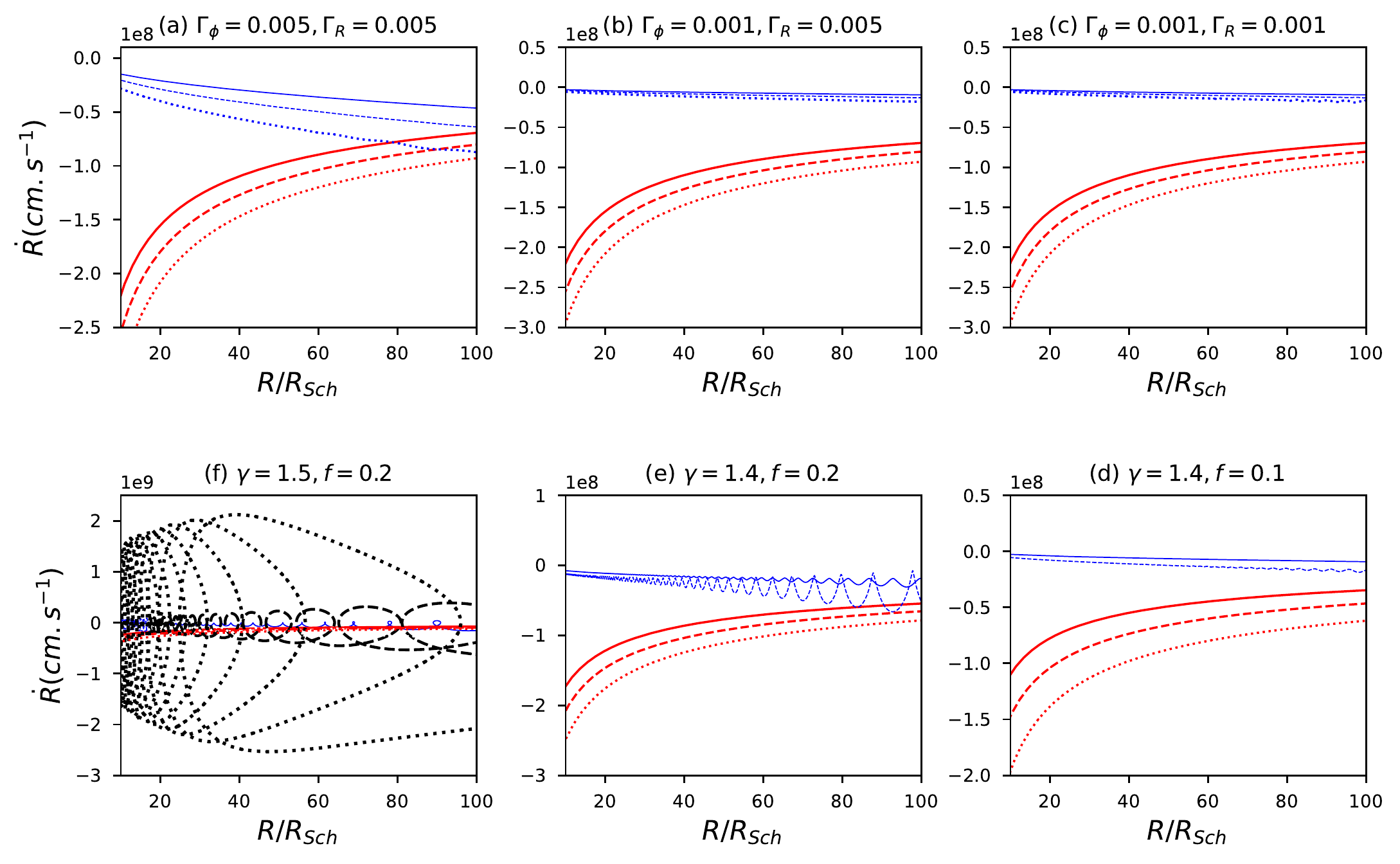}
\caption{Time-dependent variation of the radial velocity of a BLR cloud with respect to radius. Input parameters of each panel are the same as the ones in figure 7, so comparing every pair of adjacent subplots yields information about the effect of a certain parameter other than $\phi_s$. In each panel, the solid dashed and dotted lines refer to solutions with $\phi_s=0, 0.002, 0.004$, respectively. Two colors have been used in this figure; the black curves represent $\dot{R}$ and the red ones show $V_R$ of the ADAF.}
\end{figure*}

\begin{figure*}
\centering
\includegraphics[width=160mm]{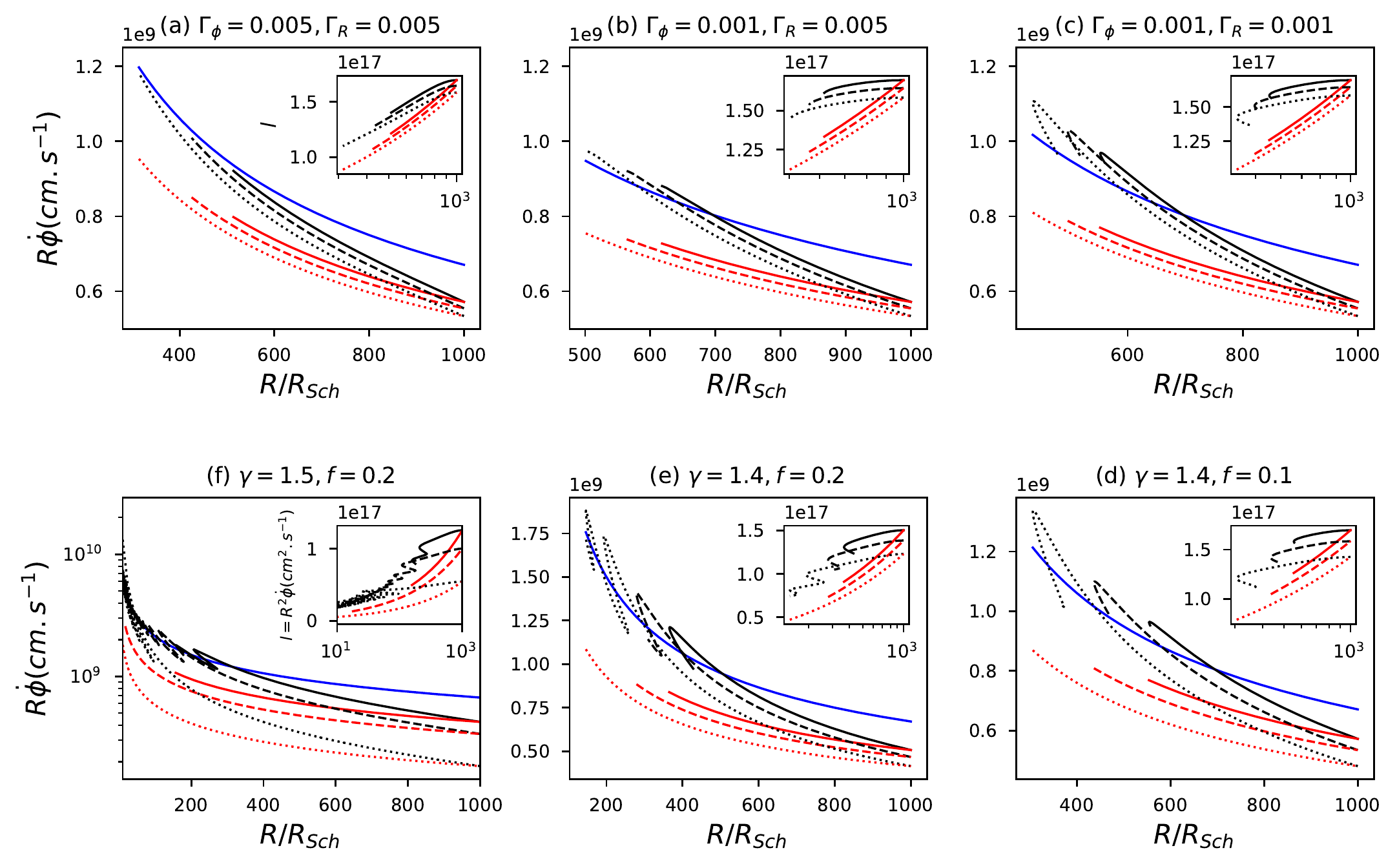}
\caption{Time-dependent variation of the rotational velocity of a BLR cloud with respect to radius.  In each panel, the solid dashed and dotted lines refer to solutions with $\phi_s=0, 0.002, 0.004$, respectively.  Similar to figure 8, the black curves represent $R\dot{\phi}$ and the red ones show $V_\phi$ of the ADAF. The set of input parameters is common to the ones in figures 7 and 8, thus we can find out the role of each parameter by comparing two related adjacent subplots. In this figure, we have also drawn plots of the angular momentum of both clumps (black curves) and ADAF (red curves) inside the main larger plots. The same line style has been used for different values of $\phi_s$ in smaller plots too.  }
\end{figure*}

\section{Thermal conduction effect on the projection of clumps}
Up to now, we have been studying semi-analytically the dynamical motion of clouds through the hot medium of ADAFs by
adopting collision-less Boltzmann equation for the ensemble of clouds. Considering a more realistic picture we may study the time-dependent orbits of such clouds as individual particles. In fact, the evaluation of the time-dependent trajectory of clumps informs us about the population properties of BLRs distributed around active galactic nuclei. Therefore, for the rest of this paper, we work on the time-dependent dynamics of a simplified single-cloud model of clumps in the central vicinity of AGN. We aim to involve the thermal conduction term in the momentum equation. ‌‌‌But for simplicity, we neglect the non-isotropic force due to the radiation of a central accretion disk. In this problem, we consider just two forces:  the gravitational force of a central black hole with mass $M$, and a drag force involve between the clump and ADAF. This drag force is in the opposite direction of the cloud and its amplitude is proportional to square its relative velocity with respect to the ambient medium, i.e. ADAF. Thus, we can write the equations of the orbital motion as follows:
\begin{equation}
{\ddot{R}}-R \dot{\phi}^2=-\frac{G M}{R^2}-f_R(\dot{R}-V_R)|\dot{R}-V_R|
\end{equation}

\begin{equation}
R {\ddot{\phi}}-2\dot{R}\dot{\phi}=-f_{\phi}(R \dot{\phi}-V_{\phi})|R \dot{\phi}-V_{\phi}|
\end{equation}
where $\dot{R}=dR/dt(=v_R)$, $\ddot{R}=d^2R/dt^2(=dv_R/dt)$, $\dot{\phi}=d\phi/dt(=v_\phi/R)$ and $\ddot{\phi}=d^2\phi/dt^2$. Like before, we have specified the velocity components of ADAF by $V_R$ and $V_\phi$ and we have also employed constant coefficients, $f_R$ and $f_\phi$ for the components of drag force.
For numerical evaluations, it seems more convenient to work with non-dimensional quantities, hence we convert Eq.(32) and (33) to the non-dimensional forms by introducing $r$ and $\tau$ as:
\[r=\frac{R}{R_0}\]
\[\tau=\frac{t}{t_0}\]
where $R_0$ and $t_0$ are the typical lengths and time scales. We can take $R_0$ equal to the radial size of the background medium of BLR, that is the outer boundary of ADAF: $R_0=1000 R_{Sch}$\footnote{ Notice that the dimensionless radius of $r$ is three orders of magnitude larger than the previous quantity of $r$ used in Eq.(23)-(25).}

Thus, the equations (32) and (33) become:
\begin{equation}
r''-r \phi'^2=-\frac{1}{r^2}-\Gamma_R\frac{R_0}{R_{Sch}}\bigg(r'-\frac{c_R}{ \sqrt{r}}\bigg) \bigg|r'-\frac{c_R}{ \sqrt{r}}\bigg|
\end{equation}
\begin{equation}
r\phi''-2r'\phi'=-\Gamma_{\phi}\frac{R_0}{R_{Sch}}\bigg(r\phi'-\frac{c_{\phi}}{\sqrt{r}}\bigg)\bigg|r\phi'-\frac{c_{\phi}}{\sqrt{r}}\bigg|
\end{equation}
where $r'=dr/d\tau, r''=d^2r/d\tau^2$, $\phi'=d\phi/d\tau$ and  $\phi''=d^2\phi/d\tau^2$, the two constants of $c_R$ and $c_\phi$ refer to the radial and azimuthal components of the velocity vector of ADAF. Also in the two above equations, we have removed the factor of $GM$ in the gravitational term by choosing $t_0=\sqrt{R_0^3/GM}$. Moreover, we have substituted $\Gamma_R/R_{Sch}$ and $\Gamma_\phi/R_{Sch}$ instead of $f_R$ and $f_\phi$.
Now we need to specify appropriate initial conditions for solving numerically the last two equations. In the absence of radiation force from the central object and neglecting the vertical component of the velocity vector of the ambient medium, i.e. ADAF, i.e. $v_z=0$, we may expect the plane of motion to be fixed at the equatorial plane. However, under the influence of drag force, the cloud's angular momentum cannot remain constant (during flight time). So we expect a spiral trajectory for the cloud's motion.

Regarding boundary conditions, we assume that our single cloud begins its journey from the initial point with these dimensionless coordinates: $(r= 1,\phi=0)$ with an initial velocity vector equals to ADAF's one, i.e. $(r'=c_R, \phi'=c_\phi)$. Knowing the initial conditions for $r,\phi,r'$ and $\phi'$, we can solve Eq.(34),(35) numerically with respect to the dimensionless time, $\tau$, and obtain $r(\tau), \phi(\tau), r'(\tau)$ and $\phi'(\tau)$. In figure 7, we have drawn the orbit of a sample cloud at the equatorial plane for three values of $\phi_s$. The black dot-dashed, blue dashed and red solid lines correspond to $\phi_s = 0, 0.002$ and $0.004$, respectively and the other input parameters for panels (a)-(c) are $\alpha=0.2$, $f=0.1$ and $\gamma=1.4$; for panels (d)-(f): $\alpha=0.1, \Gamma_\phi=\Gamma_R=0.001$. Regarding the two remained parameters, one of them is variable and the other one is common for two adjacent panels. The label of $\tau_f$ in each panel (exception for panel f) is a dimensionless time that the BLR cloud needs for one rotation around the center in the case of $\phi_s=0$. For panel (f), this duration is too long for the cloud in a medium with $\phi_s=0.004$ since during a smaller time it arrives at the capture radius, hence in the case (i.e. $\phi_s=0.004$), we have used the total flight time of the cloud, i.e. $\tau_T(=1.56)$  instead of $\tau_f(=2.03)$.

As it is seen in all panels of this figure, the motion of the cloud is a spiral around the center and this common feature appears due to the influence of drag force in this problem. Moreover, we see in all panels, that thermal conduction apparently causes the clump to rotate and accrete faster. In Fig.8, we have adopted the same input parameters and for a better and easier comparison we have chosen the same time flight for all cases (except for a gray line of panel c) and separated curves with equal values for $\phi_s$. The effect of parameters other than $\phi_s$ can be studied more conveniently in this figure. By variation of $\phi_s$, the most change in the pattern of spiral motion happens for curves with different $\alpha$, because we see a green dotted curve (with $\alpha=0.2$) in panel (a) overlaps with a red dot-dashed one (with $\alpha=0.2$) whereas, in panel (b) and (c) when $\phi_s$ is not zero, these two curves remarkably different from each other. Moreover, the panels of this figure declare that time-averaged rotational velocity ($\bar{r\phi'}$) of BLR clumps is greater with larger values of $\Gamma_\phi$,  $f$ and $\gamma$ but smaller $\Gamma_R$ and $\alpha$. Similar behavior can be deduced for the time-averaged radial velocity, i.e. $\bar{r'}$, too (except for the cases with different $\alpha$'s without thermal conduction as we mentioned before).

It can be interesting to know about the treatment of velocity components of BLR clouds during their flight time and compare it with the pattern of the velocity field of clumps in the first part of this paper. On the other hand, the estimation of BLR's velocity is vital for calculating emission-line width (FWHM), thus we have presented the instantaneous velocity of our typical cold cloud in figures 9, 10, and 11. Each panel of these three figures is corresponding to one in Fig.7. We have also added the velocity components of ADAF (marked by red color) to these panels and also the Keplerian velocity (blue) in Fig.11. Notice that Fig.9 and 10 both illustrate the radial velocity but for two regions; outer part,  $100<R/R_{Sch}<1000$ in figure 9 and the inner part, $10<R/R_{Sch}<100$ in figure 10. The solutions in the duration of $0\leq \tau\leq 2.5$ are specified by black color and beyond this time are drawn by blue color in Fig.9 and 10 (see Tabel.1 for the total flight time of each case).  As clearly seen, these figures are very different from those in Fig. 5 and 6, unlike clumps in steady-state condition, here clouds can have a positive radial velocity in some parts of their eventual path towards the center (ref. Fig.9c-f and Fig.10f). In the absence of the drag force, the orbital motion of clouds would be Keplerian and they could passes a point many times but here they cross a point up to twice (and three times in the case of panel f) but not with the same velocity or even with the same direction (look at panels c-f of Fig.9). Comparing red and black curves (of Fig.9) implies that the radial velocity of clouds at outer part may grow about one order of magnitude larger than ADAFs in the middle of their path but their rotation does not show this much difference (according to Fig.11).  Very different behavior of radial velocity is seen in Fig.10 (exception for the case of panels e, f), here there are not that many fluctuations in the magnitude of $\dot{R}$ that was in outer part. Moreover, clouds pass much more slowly in the inner part of a hot background medium with $\gamma=1.4$ and $f=0.1$.   The small plots inside the main large plots of Fig.11 display the angular momentum of clouds with respect to the radius (red color curves show the ADAF's angular momentum). According to these curves, we can see like each element of hot gas, the angular momentum of individual clumps decreases towards the center, thus they can participate in the transportation of the total angular momentum of the disk outwards and consequently cause enhancing the accretion rate effectively.

 The effect of thermal conduction on the clouds' and ADAF's velocity is apparently similar as seen in Fig.9-11. Nevertheless, we should remember that time is important for clouds. For instance, if we choose a certain time and at that time we compare the rotational velocity of two clouds embedded in mediums with different $\phi_s$ and crossing two points (with different spatial coordinates), that one with larger $\phi_s$ has a larger $R\dot{\phi}$ (look at Fig.11, compare the inner last point of each curve with that of other curves); this point can be understood better by looking at Fig.12.  In figure 12, we have shown the variation of the two components of velocity with respect to the flight time of clumps with applying the same sets of input parameters as Fig.8. Here again we can see how the presence of thermal conduction changes the other parameters effect on the velocity of BLR clouds during a certain time.
 
 
\begin{figure*}
\centering
\includegraphics[width=170mm]{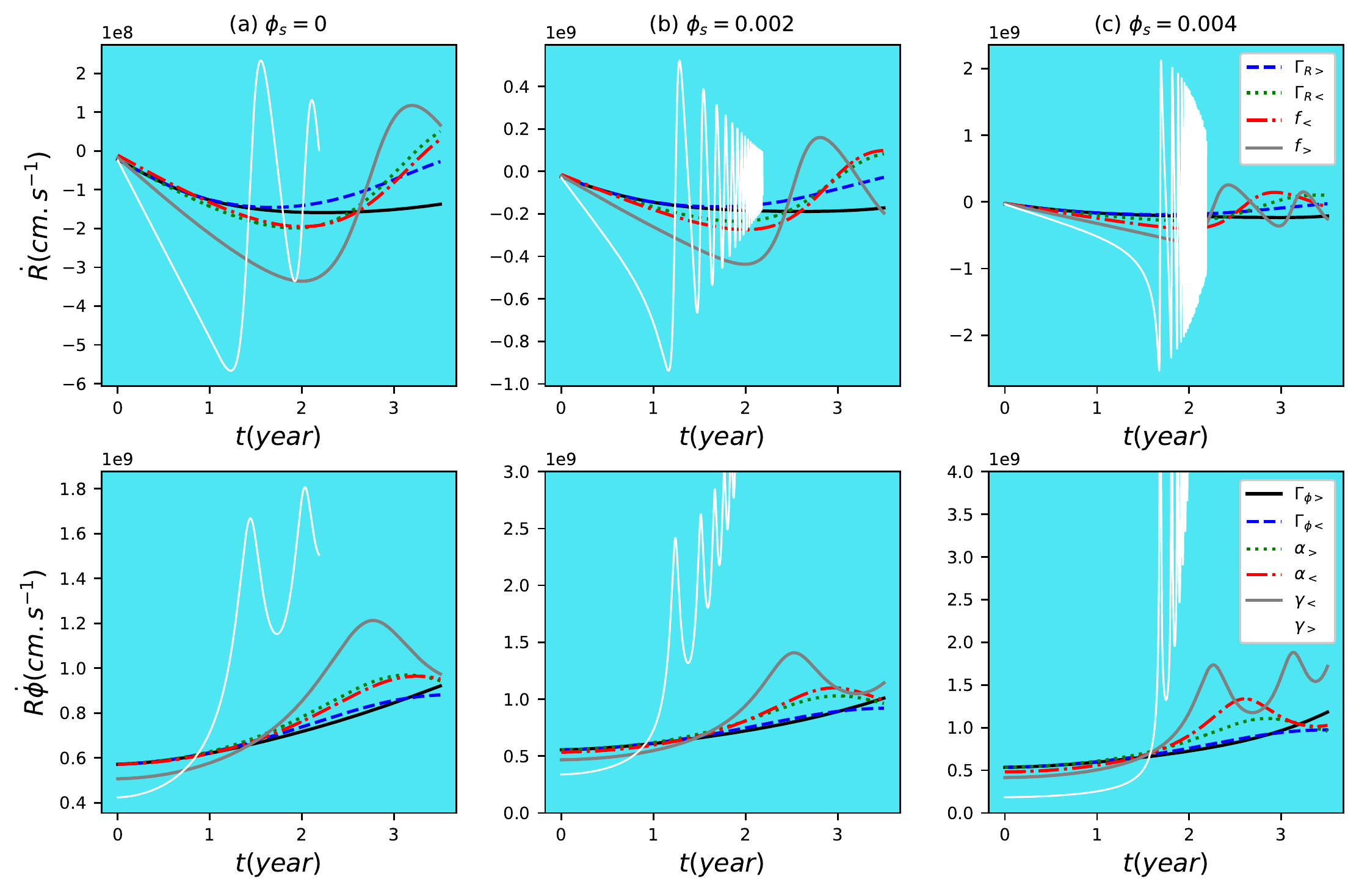}
\caption{Variation of velocity components of a BLR cloud with respect to the flight time. The setting of input parameters is the same as in figure 8. }
\end{figure*}


In Fig.13 we have calculated the total flight time ($T_T$)  of a BLR cloud as a function of the thermal conduction parameter, $\phi_s$. The time scale of this figure  has been evaluated by assuming $r_0=1000R_{Sch}\approx 0.01 pc$, $M=10^8 M_\odot$: $t_0\approx 4.42\times 10^7 s$ which is about $511.5$ days or $1.4$ years. Black color curves refer to clumps in different mediums and also red curves show the total time needed for an element of gas moves between outer and inner boundaries of ADAF, i.e.
\[T_A=-\int_{R_{in}}^{R_{out}} \frac{dR}{V_R}=\frac{2}{3 c_R}\bigg[1-\big(\frac{R_{in}}{R_{out}}\big)^{3/2}\bigg]t_0\]
 where $t_0=\sqrt{R_{out}^3/GM}$. In panel (a), we have plotted the total flight time for four different sets of $\Gamma_\phi, \Gamma_R$ and $\alpha$ with fixed values for $\gamma(=1.4)$ and $f(=0.1)$. As we know with larger $\Gamma_R$ and $\Gamma_\phi$ clumps experience stronger drag force but evidently the same variation in their magnitude cause different change in $T_T$; apparently in dynamics of BLRs $\Gamma_\phi$  acts more effective than $\Gamma_R$ (look at Fig.14). Regarding the role of $\phi_s$ in the stability of clumpy BLR systems, we see in both panels of figure 13, increasing $\phi_s$ has a negative effect on $T_T$ and makes it shorter. On the other hand, decreasing $f$ and $\gamma$ result shorter lifetime for clouds. Now, if we compare red curves with black ones, we can find out those clumps with smaller $\Gamma_\phi$ (black dashed and dotted curves in panel a) can be as stable as or even more stable than their host medium (or background ADAF).

 In order to clarify the role of each component of drag force separately, in figure 14 we have drawn the projection of a typical BLR cloud for three different cases:  a drag force  1. with two zero components; $\Gamma_R=\Gamma_\phi=0$ (black dot-dashed line), 2. with one zero component at the azimuthal direction; $\Gamma_\phi=0$ and 3. with one zero component at the radial direction; $\Gamma_R=0$. As we expect the first curve without drag force illustrates a Kepler orbit with an elliptic shape. On the other hand, although the second curve is at first elliptical, after a few moments it converts to spherical under the influence of drag force in the radial direction, thus it becomes as stable as the first case. In contrast with first and second cases, the motion of the clump deviates significantly from Keplerian in the light of a drag force with a single component in $\phi-$direction.


\begin{figure*}
\centering
\includegraphics[width=150mm]{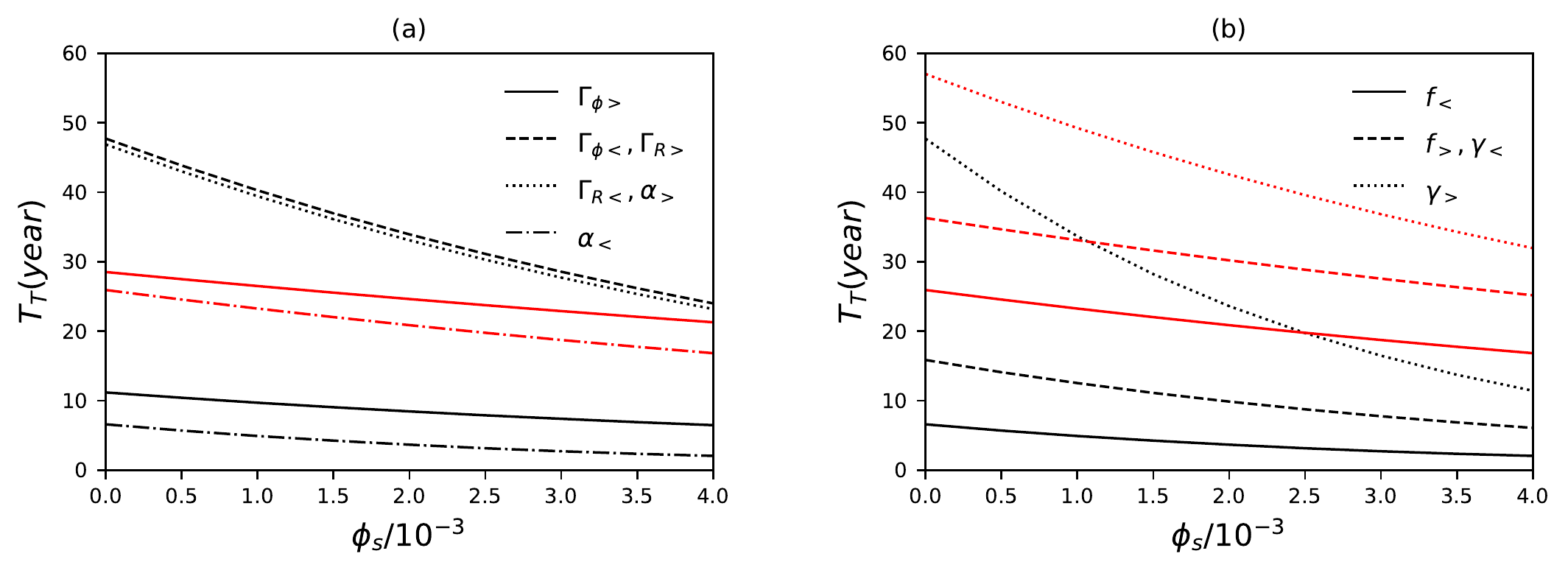}
\caption{ The total flight time of a typical BLR cloud as a function of the thermal conduction parameter with a different set of other input parameters. In panel (a), we have examined two different values of the azimutial coefficient of drag force: $\Gamma_\phi=0.005$ (solid line, labeled by $\Gamma_{\phi>}$), $\Gamma_\phi=0.001$ (dashed line, $\Gamma_{\phi<}$) both presented in black color and other parameters are the same as panels (a) and (b) of Fig.7. The other curves in panels (a) and (b) here have been drawn to study the effect of: 1. $\Gamma_R$ by trying $\Gamma_{R>}=0.005$, (presented by dashed line in panel a, other parameters are the same as Fig.7b); $\Gamma_{R<}=0.001$ (dotted line in panel a, others equal to ones in Fig.7c); 2. $\alpha$ with choosing $\alpha_>=0.2$ (dotted) and $\alpha_>=0.1$ (dot-dashed; others like Fig.7d); 3. $f$ with two values of $f_<=0.1$ (solid line in panel b; others like Fig.7d) and $f_>=0.2$ (dashed line in panel b; others like Fig.7e); 4. $\gamma$ with trying $\gamma_<=1.4$ (dashed) and $\gamma_>=1.5$ (dotted line in panel b; other parameters are the same as Fig.7f). In this figure, the curves in red color represent the total dimensionless time required for an element in ADAF reach the inner boundary (i.e. $10 R_{Sch}$) from the outer boundary ($1000 R_{Sch}$). In panel (a), for all values of $\Gamma_R$ and $\Gamma_\phi$, $V_R$ does not change, hence just red solid line is corresponding to all three set of input parameters: ($\Gamma_\phi=\Gamma_R=0.005$), ($\Gamma_\phi=0.001,\Gamma_R=0.005$) and ($\Gamma_\phi=\Gamma_R=0.001$). }
\end{figure*}
\begin{table}
	\caption{The dimensionless total flight time ($\tau_T$) of a BLR cloud with:
$\Gamma_{R,\phi}=0.001,0.005, f=0.1,0.2, \alpha=0.1,0.2, \gamma=1.4,1.5$ }
	\centering
	\begin{tabular}{l c c rrr}
		\hline\hline
		$\phi_s$  &$0$ & $0.002$& $0.004$ \\[0.5ex]
		\hline
		$\tau_T(\Gamma_{\phi>})$  &$8.41$ & $6.41$& $4.95$ \\[0.5ex]
			\hline
		$\tau_T(\Gamma_{\phi<},\Gamma_{R>})$  &$35.33$ & $25.11$& $17.77$ \\[0.5ex]
			\hline
		$\tau_T(\Gamma_{R<},\alpha_>)$  &$35.05$ & $24.86$& $17.52$ \\[0.5ex]
			\hline
		$\tau_T(\alpha_<, f_<)$  &$35.28$ & $17.69$& $8.73$ \\[0.5ex]
	 			\hline
		$\tau_T(f_>,\gamma_<)$  &$11.99$ & $7.56$& $4.73$ \\[0.5ex]
			\hline
		$\tau_T(\gamma_>)$  &$5.11$ & $2.872$& $1.565$ \\[0.5ex]
 		\hline
	\end{tabular}
	\label{tab:PPer}
\end{table}
\begin{figure}
\centering
\includegraphics[width=70mm]{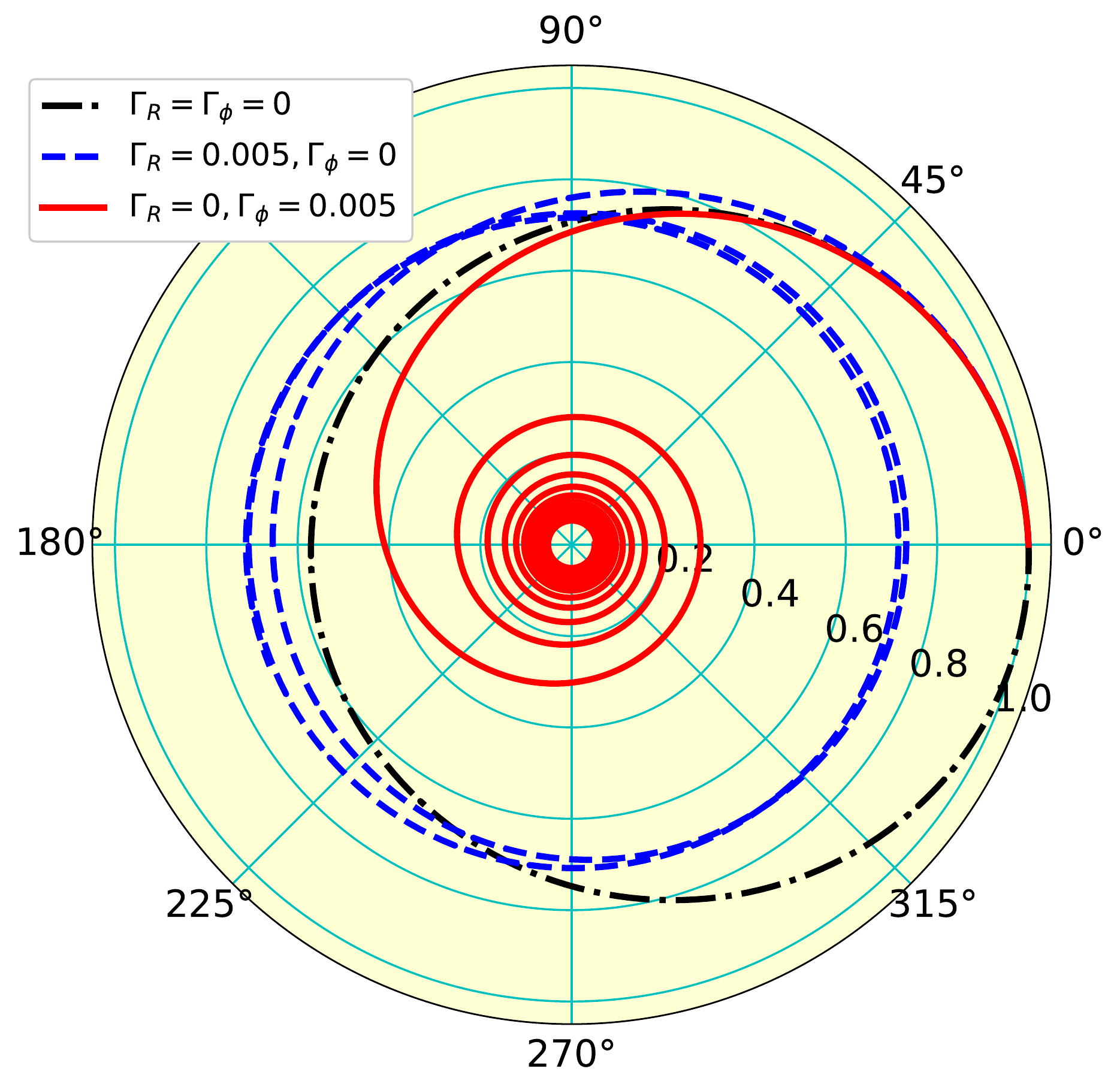}
\caption{The trajectory of a BLR cloud with three different drag forces. The black dot-dashed line corresponds to f$\Gamma_R=\Gamma_\phi=0$, the blue dashed line is related to  $\Gamma_R=0.005, \Gamma_\phi=0$ and red solid line is evaluated by choosing $\Gamma_R=0, \Gamma_\phi=0.005$. The other parameters are: $\alpha=0.2, f=0.1$ and $\gamma=1.4$. The used fight times in this figure are $\tau_f=5, 10$ and $5.5$ for the cases with $\Gamma_{R,\phi}=0$,  $\Gamma_\phi=0$ and $\Gamma_R=0$, respectively.   }
\end{figure}

\begingroup

 \section{Summary and conclusion}
In this paper, we investigated the effect of thermal conduction on the dynamics of cold clouds (or clumps) embedded in a hot accretion flow (ADAF) which experiences a drag force. W‌‌‌e took two different approaches to this problem: firstly, we supposed the clouds behave like collision-less particles and move with a velocity vector whose averaged value equals their host medium velocity, and secondly, we concentrated on clouds individually to find the position and instantaneous velocity of a typical cloud. In all parts of this study, we presumed axisymmetric and steady-state configuration for the background medium of clumps as long as we supposed that all particles have negligible vertical motions. The effect of thermal conduction appeared in this problem due to involving drag force between clouds and ADAF. We employed ADAF solutions including thermal conduction presented by Tanaka \& Menou (2006). For the first part, we followed Wang, Cheng \& Li (2012) and applied Boltzmann equations in order to obtain the root mean velocity square of the clumps (or two components of velocity dispersion), i.e. $\langle v_R^2\rangle^{1/2}$ and $\langle v_\phi^2\rangle^{1/2}$. The value of $\langle v_R^2\rangle^{1/2}$ is important since the capture rate of the clumps at the inner edge of
the disc is directly proportional to the ratio of $\langle v_R^2\rangle^{1/2}/\langle v_R\rangle$ (ref. Eq.34 of WCL12). We should mention that our results were a bit different even in the absence of thermal conduction with ones presented by WCL12 because of the opposite direction of our drag force; we ratiocinated that after applying the approximation of equal average velocity with ADAF's one, an average value of drag force must remain at the same sign as majority particles drag force sign (in the opposite direction of their relative velocity with ADAF). This tiny change in the drag force formula led to clumps with $\langle v_\phi^2\rangle^{1/2}$ smaller than the rotational velocity of ADAF that was evaluated as larger in the previous work. Furthermore, our adopted drag force had an effect on a critical (or minimum) value for the drag force coefficient in the azimuthal direction instead of the radial direction in WCL12.


For the second part, we found the trajectory of a single clump in the same way as a classical two-body problem.  The similar results for the influence of thermal conduction were achieved from both parts; when we added a larger permissible (ref. Fig.4) value for the thermal conduction parameter, $\phi_s$, the solutions changed to reduce velocity in the azimuthal direction but growing it in the radial direction. However, we found out the function of rotational velocity of a cloud behaves in opposite way with time; it may become larger with a greater $\phi_s$ (ref. Fig.12). Furthermore, we compared the total flight time (needed a clump to move from outer region of ADAF to the capture radius) of BLR clouds with the total dynamical time of a volume element of ADAF. We faced to this point: if clouds move with a very small or even zero friction in the azimuthal direction, they can survive from the central captured region of ADAF. On the other hand, we saw the existence of a little friction along $\phi-$direction modifies a Kepler orbit of clumps to a spiral one and can make it a short-lived object. The similar works had been done by Shadmehri (2015, who studied clouds in a laminar flow) and Khajenabi (2016; who worked on BLRs in a turbulent flow); both considered drag force and also repulsive radiation force from the central object. For simplicity and having the similar condition as the first part of our work, we did not take into account radiation force on this problem and assumed clumps moved at the equatorial plane under the influence of central gravity and drag force. We found out how much the stability of a BLR system can be changed by characteristic parameters (that is, $\alpha, f$ and $ \gamma$) of their hot host medium.

To improve this work, it is good for both parts to write equations in spherical coordinates and apply 2D ($r-\theta$) solutions of ADAFs which included $v_\theta$ and let us study the vertical motion of clumps too. For little improvement, we can use a formula composed of both linear and quadratic relative velocity (between the clumps and the gas) with two optional coefficients ($0,1$) set by the Reynolds number.


\section*{ACKNOWLEDGMENTS}
The authors thank the anonymous referees for the careful reading of the manuscript and their insightful and constructive comments. We hereby acknowledge the Sci-HPC center of the Ferdowsi University of Mashhad where some of this research was performed. We have made extensive use of the NASA Astrophysical Data System Abstract Service. This work was supported by the Ferdowsi University of Mashhad under grant no. 54195 (1399/12/24).

\section*{DATA AVAILABILITY STATEMENT}
No new data were generated or analyzed in support of this research

\end{document}